\documentclass[useAMS,usenatbib]{mn2e}
\usepackage{graphicx}
\usepackage{pslatex}
\usepackage{natbib}
\usepackage{txfonts}

\newcommand{\vv}[1]{\bmath{#1}}
\newcommand{\mtr}[1]{\mathbfss{#1}}
\begin{document}

\title[YORP torques with 1D thermal model]{YORP torques with 1D thermal model}

\author[S. Breiter and P. Bartczak and M. Czekaj]{S. Breiter$^{1}$\thanks{E-mail:
breiter@amu.edu.pl} and P. Bartczak$^{1}$\thanks{E-mail:
przebar@amu.edu.pl} and M. Czekaj$^{2}$\thanks{E-mail: mczekaj@am.ub.es}\\
$^{1}$Astronomical Observatory, A. Mickiewicz University, Sloneczna 36, PL60-286 Pozna\'{n}, Poland\\
$^{2}$Departament d'Astronomia i Meteorologia, Unversitat de Barcelona, Av. Diagonal 647, 08028 Barcelona, Spain}

\date{}

\pagerange{\pageref{firstpage}--\pageref{lastpage}} \pubyear{2009}

\maketitle
\label{firstpage}

   \begin{abstract}
    A numerical model of the Yarkovsky-O'Keefe-Radzievskii-Paddack (YORP) effect for objects defined in terms of a triangular mesh is described.
    The algorithm requires that each surface triangle can be handled independently, which implies the use of a 1D thermal model.
    Insolation of each triangle is determined by an optimized ray-triangle intersection search. Surface temperature is modeled with a spectral
    approach; imposing a quasi-periodic solution we replace heat conduction equation by the Helmholtz equation. Nonlinear boundary conditions are
    handled by an iterative, FFT based solver. The results resolve the question of the YORP effect in rotation rate independence on conductivity
    within the nonlinear 1D thermal model regardless of the accuracy issues and homogeneity assumptions. A seasonal YORP effect in attitude is revealed
    for objects moving on elliptic orbits when a nonlinear thermal model is used.
   \end{abstract}

\section{Introduction}

Referring to partial results of his predecessors (most notably \citet{Pad:69}), \citet{Rub:00} forged the acronym `YORP effect' (Yarkovsky-O'Keefe-Radzievskii-Paddack) to describe the influence
of radiation effects on the rotation of a Sun-orbiting object. The radiation incident on the surface of a celestial body acts in three different ways:
by the direct pressure, by the recoil force of reflected photons, and by the thermal radiation force.

According to a simple geometric argument of \citet{Rub:00}, further elaborated  by \citet{NV:08a} and \citet{RP:2010}, the average torque due to direct radiation pressure vanishes.
Physical properties of asteroid surfaces do not suggest a significant contribution of specular reflection,
so we may focus on the remaining two phenomena: scattered (i.e. diffusively reflected) radiation
and thermal re-radiation, defining the YORP torque $\vv{M}$ as the sum
\begin{equation}\label{M:full}
    \vv{M} = \vv{M}_\mathrm{d} + \vv{M}_\mathrm{t},
\end{equation}
of the torque $\vv{M}_\mathrm{d}$ generated by the  scattered radiation, and of
the grey body thermal radiation torque $\vv{M}_\mathrm{t}$.

Within our present Lambertian model, the primary definitions of the YORP torque components
in reference frame attached to an object's centre of mass  are given as integrals over the body surface:
\begin{equation}\label{M:d}
    \vv{M}_\mathrm{d} = - \,\frac{2}{3\,c}
 \,\oint_S  A\,E\,\vv{r} \times \mathrm{d}\vv{S},
\end{equation}
where $A\,E$ is the scattered fraction of incident power flux -- the product of albedo $A$ and of the flux $E$
hitting the infinitesimal surface element $\mathrm{d}S$, and
\begin{equation}\label{M:t}
    \vv{M}_\mathrm{t} = - \,\frac{2\,\sigma}{3\,c}
 \,\oint_S  \epsilon_\mathrm{t}\,T^4\,\vv{r} \times \mathrm{d}\vv{S},
\end{equation}
where the re-radiated energy flux is the product of the Stefan-Boltzmann constant
$\sigma$, grey body emissivity factor $\epsilon_\mathrm{t}$ and the fourth power of surface temperature $T$.
Both terms are divided by the velocity of light $c$.

Conservation of energy on the body surface implies that
\begin{equation}\label{bc:1}
 \epsilon_\mathrm{t} \, \sigma \, T^4 + K\,   \vv{n} \cdot \nabla T - (1-A)\,E = 0,
\end{equation}
so the absorbed flux $(1-A)\,E$ is distributed between the re-radiation, proportional to $T^4$, and conduction
term given as the product of thermal conductivity $K$ and the normal derivative of temperature -- the gradient
projected on the outward normal unit vector $\vv{n}$.

Substituting the boundary condition (\ref{bc:1}) into Eq.~(\ref{M:t}) we can merge a part of $\vv{M}_\mathrm{t}$ with
$\vv{M}_\mathrm{d}$, so that the YORP torque becomes the sum
\begin{equation}\label{M}
    \vv{M} = \vv{M}_\mathrm{R} + \vv{M}_\mathrm{c},
\end{equation}
of the principal term
\begin{equation}\label{M:R}
    \vv{M}_\mathrm{R} = - \,\frac{2}{3\,c} \,\oint_S  E\,\vv{r} \times \mathrm{d}\vv{S},
\end{equation}
and the complement due to conductivity
\begin{equation}\label{M:c}
    \vv{M}_\mathrm{c} =  \,\frac{2}{3\,c} \,\oint_S  K\, \left( \vv{n} \cdot \nabla T \right)\,\vv{r} \times \mathrm{d}\vv{S}.
\end{equation}

The subscript $R$ refers to the usual `Rubincam approximation' of zero conductivity, and the problem of finding $\vv{M}_\mathrm{R}$
is actually an exercise in computational geometry. Its most difficult part is the evaluation of $E$, discussed in Sect.~\ref{Insol}.
Computing the conductivity term $M_\mathrm{c}$ requires solving the heat diffusion equation. Our simplified 1D thermal model is presented
in Sect.~\ref{Thermal}. Two of its possible extensions are given in Appendices \ref{Regol} and \ref{Finde}, but the latter serves mostly
as a theoretical argument and has not been implemented.
The results of test runs with asteroids $\mathrm{1998~KY_{26}}$ and 6489 Golevka are presented in Sect.~\ref{tests}.
In our opinion, they reveal a previously unnoticed seasonal YORP effect in attitude.
Additional assumptions of our model are enumerated in Sect.~\ref{Prelim}, but we hope to relax them in future.

\section{Preliminaries}
\label{Prelim}

\subsection{Body shape model}

Although there are many possible ways to describe the shape of a celestial body, the YORP studies practically rely on two variants:
a spherical harmonics model (typical for analytical considerations) or a triangular mesh. We adopt the latter as more general, capable
of representing even very irregular objects, and more suitable for the occlusion tests. As a consequence, an integral over the body
surface becomes the sum of cubatures over all triangular patches forming the mesh. Of course, the real information about the
surface points is given only at the vertices $\vv{r}_i$, so the values of distance or any other coordinates dependent function
have to be interpolated on a patch. In principle, it can be done using various interpolation rules, even the ones that involve the whole set
of vertices, but in the YORP practice all authors rely on the local, linear interpolation, considering flat triangles and replacing
all cubatures over triangular patches by the first order Gaussian midpoint rule
\begin{equation}\label{cub}
    \int_{S_j} f(\vv{r})\,\mathrm{d}S_j \approx  f(\vv{r}_j)\,S_j,
\end{equation}
where $S_j$ is the area of a triangle determined by vertices $\vv{r}_0^j$, $\vv{r}_1^j$, $\vv{r}_2^j$, and $\vv{r}_j$ is the centroid
\begin{equation}\label{centr}
    \vv{r}_j = \frac{1}{3}\,\left( \vv{r}_0^j + \vv{r}_1^j + \vv{r}_2^j \right).
\end{equation}
In particular, the oriented surface vectors $\vv{S}_{j} = \vv{n}_j\,S_j$ are constant on each triangular face, easily computed as
\begin{equation}\label{Sj}
  \vv{S}_j = \frac{1}{2}\,\left(\vv{r}_1^j - \vv{r}_0^j\right) \times \left(\vv{r}_2^j - \vv{r}_0^j \right).
\end{equation}

Of course, the mesh should be properly oriented, so that $\vv{S}_j$ computed from Eq.~(\ref{Sj}) is always directed along the outward normal. The routine tests
rely on checking the Gauss identity
\begin{equation}\label{Gauss}
    \sum_{j=1}^{N_\mathrm{m}} \vv{S}_j = \vv{0},
\end{equation}
followed by asking if the volume resulting from the sum of oriented tetrahedral simplices
\begin{equation}\label{vol}
   V = \frac{1}{6} \sum_{j=1}^{N_\mathrm{m}} \vv{r}_0^j \cdot \left( \vv{r}^j_1 \times \vv{r}^j_2 \right),
\end{equation}
is positive, when all $N_\mathrm{m}$ faces are included. Yet, even if both tests have been passed, there remains a number of possible degeneracies,
like edges shared by more than two triangles, duplicated vertices etc., that are best to be checked before using the mesh.

Thus, for a model of a celestial body with $N_\mathrm{m}$ triangular faces, the YORP torque is approximated as a sum
\begin{equation}\label{sumtr}
    \vv{M} = \sum_{j = 1}^{N_\mathrm{m}} \vv{M}^j,
\end{equation}
with
\begin{equation} \label{M:dsc}
     \vv{M}^j = -\frac{2}{3\,c} \,\left( E_j + Q_j \right)\,\vv{r}_j \times \vv{S}_j ,
\end{equation}
where two terms in the bracket are: $E_j$ -- the incident power flux evaluated at the centroid $\vv{r}_j$,
and
\begin{equation}\label{Tprim}
 Q_j =  - K \, \vv{n}_j \cdot\left[\nabla T \right]_{\vv{r}_j}.
\end{equation}
These two terms are responsible for the Rubincam part and the conductivity complement, respectively.

\begin{figure}
 \begin{center}
  \includegraphics[width=8cm]{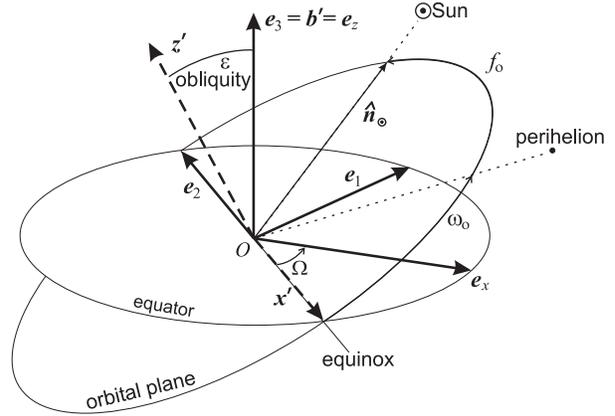}
  \end{center}
  \caption{Reference frames and angles in a body-centered system. }
  \label{fig:01}
\end{figure}

\subsection{Dynamics}
\label{dynam}

Although recent works of \citet{VBNB:07} and \citet{CS:09} have revealed the importance of tumbling rotation for the dynamics under the YORP torque,
we adhere to the usual assumption of the rotation around the principal axis of inertia -- the $\vv{e}_z$ unit vector of the body-frame basis.
The principal axis mode remains a decent approximation over significant fragments of the evolutionary paths presented by \citet{VBNB:07},
and in some instances it may be possible to incorporate tumbling by rotating the basis with respect to the principal axes.
So, we consider the principal axis mode equations of motion for the rotation rate $\omega$, the obliquity $\varepsilon$ (the angle between the normal to the orbital plane and
the spin vector $\vv{\omega}$, parallel to $\vv{e}_z$), and the sidereal time $\Omega$ (measured from the ascending node of the
Sun on the object's equator to the body-frame basis $\vv{e}_x$  vector)
\begin{eqnarray}
  \dot{\omega} &=&  \frac{\vv{M} \cdot  \vv{e}_3 }{C}, \label{eqm:1} \\
  \dot{\varepsilon} & = & \frac{\vv{M}\cdot  \vv{e}_1}{
  \omega\,C}, \label{eqm:2} \\
  \dot{\Omega} & = &
 \omega - \frac{\vv{M}\cdot  \vv{e}_2}{
  \omega\,C\,\tan{\varepsilon}}, \label{eqm:3}
\end{eqnarray}
where $C$ designates the maximum moment of inertia in the principal axes frame.

According to the above equations, the dynamics is governed by the components of the YORP torque $\vv{M}$ in
another kind of equatorial reference frame $(\vv{e}_1,\,\vv{e}_2,\,\vv{e}_3)$ (see Fig.~\ref{fig:01}) with the same (centre of mass) origin as the body frame
$(\vv{e}_x,\,\vv{e}_y,\,\vv{e}_z)$,
with the same $z$ direction, but with the remaining axes related to the equinox instead of to the principal axes
\citep{BVN:09}
\begin{equation}\label{vecs}
\begin{array}{l}
 \vv{e}_1  =
   \sin{\Omega}\,\vv{ e}_x
   +\cos{\Omega}\,\vv{ e}_y ,\\
 \vv{e}_2   =
   - \cos{\Omega}\,\vv{ e}_x + \sin{\Omega}\,\vv{ e}_y, \\
     \vv{e}_3  =  \vv{ e}_z. \\
\end{array}
\end{equation}

\subsection{Average YORP effect}

From the point of view of long term, systematic influence of the YORP effect, we are mostly interested in the mean values
of $\dot{\omega}$, $\dot{\varepsilon}$ and $\dot{\Omega}$, with all daily and orbital periodic effects averaged out.
Thus, assuming a uniform rotation with constant frequency $\omega$ and the Keplerian motion around the Sun, we need
to find the mean values, defined for any function $f$
\begin{equation}\label{mv}
    \langle  f \rangle = \frac{1}{4\pi^2} \int_0^{2\pi} \int_0^{2\pi} f \,\mathrm{d}\ell\,\mathrm{d}\Omega,
\end{equation}
where $\ell$ is the mean anomaly of the Sun.
We perform this averaging indirectly, working with discrete Fourier transforms (DFT) of $f(\ell,\Omega)$.
Using the conventions described in Appendix~\ref{ap:A}, we assume
\begin{equation}\label{mv:1}
    \langle f \rangle =   \vv{\hat{f}}[0],
\end{equation}
where $\vv{\hat{f}} = \mtr{F}_2\,\vv{f}$ is the DFT of the function $f$, and $\vv{f}$ is the vector of samples of this function
at different $\Omega$ and $\ell$ values.

According to Eq.~(\ref{M:dsc}),  the mean value of $\vv{M} \cdot \vv{e}_3$ on a single triangle $j$ is simply
\begin{equation}\label{M3:a1}
  \left\langle  \vv{M}^j \cdot \vv{e}_3 \right\rangle = -\frac{2}{3\,c} \,\left( \vv{\hat{E}}_j[0] + \vv{\hat{Q}}_j[0] \right)\,\left(\vv{r}_j \times \vv{S}_j\right) \cdot \vv{e}_z,
\end{equation}
where $\vv{S}_j$ is computed in the body frame, so the scalar product with $\vv{e}_3 = \vv{e}_z$ amounts to selecting the third component.
But the averaging of the next two terms is slightly more involved: confronting Eq.~(\ref{vecs}) with (\ref{trig}) we find
\begin{eqnarray}
  \left\langle  \vv{M}^j \cdot \vv{e}_1 \right\rangle & = &
    - \frac{2}{3\,c} \,\left[ \Re\left( \vv{\hat{E}}_j[1] + \vv{\hat{Q}}_j[1] \right)\,\left(\vv{r}_j \times \vv{S}_j \right) \cdot \vv{e}_y  \right. \nonumber \\
    & & \left. - \Im\left( \vv{\hat{E}}_j[1] + \vv{\hat{Q}}_j[1] \right)\,\left(\vv{r}_j \times \vv{S}_j \right) \cdot \vv{e}_x \right] ,
 \label{M1:a1} \\
      \left\langle  \vv{M}^j \cdot \vv{e}_2 \right\rangle & = &
     \frac{2}{3\,c} \,\left[ \Re\left( \vv{\hat{E}}_j[1] + \vv{\hat{Q}}_j[1] \right)\,\left(\vv{r}_j \times \vv{S}_j \right) \cdot \vv{e}_x \right. \nonumber \\
    & & \left. + \Im\left( \vv{\hat{E}}_j[1] + \vv{\hat{Q}}_j[1] \right)\,\left(\vv{r}_j \times \vv{S}_j \right) \cdot \vv{e}_y \right].
 \label{M2:a1}
\end{eqnarray}
The above expressions make use of the Hermitian property of the DFT for a real-valued function, and assume that in the sampling described in Appendix~\ref{ap:A}
the first angle is $\phi = \ell$, and the second is $\psi = \Omega$.

\section{Rubincam terms}
\label{Insol}

\subsection{Insolation function}

The major difficulty in dealing with the Rubincam part of the YORP effect is the computation of $E$, known as the insolation
or irradiation function. In principle, the incident energy flux hitting a given surface element is a sum of two
components: the direct flux from the Sun, and the radiation exchange complement, i.e. the energy coming from other elements
of the body surface (either reflected or re-emitted). In the present paper we adhere to the approximation used in all previous
works and consider only the direct part
\begin{equation} \label{E:first}
E  =  \Phi\,\xi(\vv{r},\,\vv{\hat{n}},\,\vv{\hat{n}}_\odot) \,\vv{\hat{n}}\cdot\vv{\hat{n}}_\odot,
\end{equation}
where $\Phi$ designates the solar radiation power flux at a given distance of the body from the Sun $r_\mathrm{o}$.
Using the solar constant
$\Phi_0 \approx 1366~\mathrm{W\, m^{-2}}$, we have
\begin{equation}\label{flux}
    \Phi = \Phi_0\,\left(\frac{d_0}{r_\mathrm{o}}\right)^2,
\end{equation}
with the reference distance  $d_0 = 1\,\mathrm{au}$.

Defining and computing the visibility function $\xi$ is the heart of the problem. Its values are either $\xi = 0$, when the Sun is not visible
over the current surface element, or 1 otherwise. For convex bodies, $\xi$ depends only on the scalar product of the outward normal unit vector
$\vv{\hat{n}}$ and the unit vector directed to the Sun $\vv{\hat{n}}_\odot$. In other words, whenever the zenith distance of the Sun
is less than $90\degr$, the visibility function $\xi=1$, because the formal horizon (a local tangent plane perpendicular to $\vv{\hat{n}}$) and
the actual horizon (the part of a celestial hemisphere not occluded by other surface elements) coincide for a convex object.
In this case, computing $\xi$ is so cheap and easy, that often the bodies of an arbitrary shape are formally treated as convex when
computing $E$, which is necessary in analytical theories \citep{NV:07,NV:08,Mysen:08,BM:08,BVN:09}, and handy in numerical or some semi-analytical
models \citep{VBNB:07,SG:08,CS:09}. However, the weakness of such pseudo-convex treatment for irregular, bouldered and cratered objects, testified by
\citet{SMG:08} and -- in a quite different form -- by \citet{BBCOV:09}, suggests to avoid it whenever possible, unless the shape model is known in advance to be
convex (e.g. when it comes from the convex lightcurve inversion algorithm).

Leaving aside the visibility function algorithms, to be discussed in next subsections, we begin computations with tabulating
the flux $\Phi$ and the components of $\vv{\hat{n}}_\odot$ in the orbital frame for the mean anomaly $\ell$ sampled
at $N$ equidistant points in the full angle range $0 \leqslant \ell < 2\pi$. In the orbital frame, the direction cosines
are formally
\begin{equation}\label{orb}
    \left[ \vv{\hat{n}}_\odot\right]_\mathrm{orb} =
\left(
  \begin{array}{c}
    \cos{\left(\omega_\mathrm{o}+ f_\mathrm{o}\right) } \\
    \sin{\left(\omega_\mathrm{o}+f_\mathrm{o}\right)} \\
    0 \\
  \end{array}
\right),
\end{equation}
where $\omega_\mathrm{o}$ is the argument of perihelion and $f_\mathrm{o}$ is the true anomaly of the Sun.
Thus the two nonzero components and $\Phi$ are tabulated once, before the the main loop over surface triangles begins, so
the cost of solving Kepler equation is relatively negligible. Other quantities precomputed before the main algorithm
starts are: centroid positions $\vv{r}_j$, areas $S_j$, and unit normal vectors $\vv{\hat{n}}_j$ associated with each triangular face.

Given a pair of mean anomaly and rotation phase $(\ell,\,\Omega)$ we transform the solar vector to the body frame by means
of two rotations: around the first axis by angle $(-\varepsilon)$, and then around the third axis by angle $\Omega$, so that
\begin{equation}\label{nbf}
     \vv{\hat{n}}_\odot  =
\left(
  \begin{array}{ccc}
    \cos{\Omega } & \cos{\varepsilon} \sin{\Omega} & * \\
    -\sin{\Omega } & \cos{\varepsilon} \cos{\Omega} & * \\
    0 & \sin{\varepsilon} & * \\
  \end{array}
\right)\, \left[ \vv{\hat{n}}_\odot\right]_\mathrm{orb},
\end{equation}
where `*' are placeholders for irrelevant matrix entries.

\subsection{Visibility function}

The fundamental operation in the evaluation of the visibility function is the `stabbing query', i.e. testing the intersection
of a ray
\begin{equation}\label{ray}
    \vv{w} = \vv{r}_j + \eta \, \vv{\hat{n}}_\odot, \qquad \eta > 0,
\end{equation}
with other surface triangles. This standard tool of computational geometry is well documented \citep{MT97}, so we skip the details
focusing on a less trivial question: how to minimize the number of its calls.

Obviously, there is no need to perform the query when the Sun is below the formal horizon, i.e. when $\vv{\hat{n}}_j \cdot \vv{\hat{n}}_\odot \leqslant 0$.
So, the most straightforward selection tool is to create for each $j$ a list of all triangles with at least one vertex above the formal horizon and, if
$\vv{\hat{n}}_j \cdot \vv{\hat{n}}_\odot > 0$, perform the queries only with triangles from the list. But expecting that the list should be short
is  wishful thinking, based upon a false intuition of a flat landscape with distant mountains on the horizon and plenty of clear sky above a spectator's
head.

Quite a number of difficult to spot errors may arise if an optimized visibility algorithm is created with such a picture in mind.
If there are craters or boulders on an asteroid, one should rather try to imagine the landscape seen by an ant climbing a pit or walking on a side of a boulder
surrounded by other rough terrain features. The region of a clear sky can be a small, irregular, non-convex area, and its intersection with the daily Sun path
can be a union of disjoint segments.\footnote{However prudent seems the approach described by Scheeres (2007), detecting the longitudes of Sun path intersections with facets edges, it may involve
another kind of subtleties responsible for the differences between Scheeres et al. (2008) and Breiter et al. (2009) concerning the influence of shadowing
on the YORP effect for 25 143 Itokawa.} For a triangle on a boulder or a crater side, up to 90\% of remaining triangles may stretch above the formal horizon,
and it means that the number of queries should be additionally optimized.

An optimization method, very briefly reported in the paper of \citet{Statler:09}, relies on a horizon map -- a 1D array of maximum elevation of surface features
above the formal horizon on a grid of local azimuth values for a given triangle. However, it is not clear from the author's description how far his
approach is based upon the `hills on a horizon' paradigm and whether he avoided the
problems arising when some triangles overhang the local zenith, because then the altitude of clear sky has both the lower and the upper bound
(smaller than $90\degr$).

Another, more robust way to handle the optimization, applied in papers like \citep{VC:02,CV:04,DVK:08}, and described in \citep[Appendix~B2]{Cap:08}, amounts to creating
a huge collection of 2D visibility tables for a specific object. For each surface triangle, a longitude-latitude Mercator map with 0/1 values on a $1\degr \times 1\degr$  grid
is first computed and stored in a file. During the YORP computation, the longitude and latitude of the Sun are rounded to full degrees and compared with the
related entry of the visibility table. Creating the tables is time consuming, but performed only once for a given object. The drawbacks are: fixed discretisation
error with uneven resolution on a sphere, and huge file space requirements. The largest shape model attacked with this approach was the triangulated Itokawa shape
with $196\,608$ facets \citep{DVK:08}.

Our approach is an attempt to join the robustness of \citet{Cap:08} with the potential efficiency of \citet{Sch:07} or \citet{Statler:09}.
The method has already proved its valor in computing the Rubincam part of the YORP effect for Itokawa and Eros using their models of over $3 \times 10^6$ triangular faces
\citep{BBCOV:09}. Computing the YORP torques for a number of obliquity values $\varepsilon_i$, we use the following arrangement of loops:
triangles(obliquity(orbit(rotation)). The efficiency of our approach hinges upon the possibility of considering
surface elements one by one in the outermost loop, which is possible within the assumptions of the illumination and
thermal model of the present work, although suppressing the present restrictions in future, we will most likely
find ourselves in a less comfortable situation.

For a current surface triangle $j$ we first create a `horizon array', partitioning the local hemisphere into a fixed number of sectors (along meridians) and zones (along constant altitude
circles). A typical setup uses about 100 sectors and 64 zones. Each  triangle above the formal horizon is centrally projected onto the unit sphere with the origin at $\vv{r}_j$ in order to
find its `bounding box' in azimuth and altitude. The problem has its own subtleties: the extreme azimuth values are those of the vertices, but care must be taken about
the cases of crossing the zero meridian; on the other hand, the extremes of altitude are often different from the altitude of vertices due to the bending of a straight edge
in central projection. And, last but not least, if some triangle is intersected by the local zenith line (parallel to the normal vector $\vv{\hat{n}}_j$),
it should be marked as a `zenith triangle' and requires a special treatment, having a constant altitude circle as the bounding box.

In addition to the bounding box determination, each triangle $k$ is also labeled as foreground or background object, depending on the sign of the scalar product
of its outward normal vector $\vv{\hat{n}}_k$  and the relative position of its centre $\vv{r}_k - \vv{r}_j$. Obviously, if any ray $\vv{w}$ intersects
a foreground triangle, it must intersect a background triangle as well, hence -- for economy of time and storage -- we select a less populated of the
foreground and background subsets of faces above the formal horizon as the candidates for future stabbing queries.

Once the first loop over triangles $k \neq j$ is completed, the horizon array is dynamically created with an appropriate size.
Then, in the second loop over $k \neq j$, the number $k$ is stored in the lists referring to all zone-sector cells
covered by the bounding box of the $k$-th triangle.

Thus we create the horizon array -- a set of lists containing possible occluders for a given solar altitude and azimuth.
Actually, the array covers the entire hemisphere only in the presence of a zenith triangle. If no such face is detected, we record the the
bounding altitude of the clear sky cap and set the horizon array cells subdividing only the sector between the formal horizon and the
clear sky limit circle. After that, the remaining computations are straightforward: fixing the value of obliquity (or opening the obliquity loop)
we sample the rotation phase and mean anomaly, and for each pair of these angles compute the Sun vector $\vv{\hat{n}}_\odot$. If the Sun is above the formal horizon,
we select an appropriate entry of the horizon array and perform stabbing queries with triangles from the list, until we record the intersection ($\xi = 0$) or
the end of the list is reached ($\xi = 1$). Having collected all values of the insolation function $\vv{E}_j \in \mathbb{R}^{N^2}$, we perform the DFT
and find the requested amplitudes $\vv{\hat{E}}_j[0]$, and $\vv{\hat{E}}_j[1]$. Of course, a simple arithmetic mean can be used instead of the DFT (as we did in \citet{BBCOV:09}),
but the complete spectrum is required to compute the conductivity terms, as described in the next section.

\section{Conductivity terms}
\label{Thermal}

\subsection{Plane-parallel model}
\label{sect:PPM}

The surface temperature gradient, required by the conductivity complement, is obtained by solving the heat diffusion equation
\begin{equation}\label{Four:1}
   \nabla \cdot K\,\nabla T =  \rho \, c_\mathrm{p}\, \frac{\partial T}{\partial t},
\end{equation}
where $\rho$ is the density and $c_\mathrm{p}$ is the specific heat capacity of the object.
If we assume that conductivity $K$ is independent on temperature and has the same value in the entire volume of the
body, we reduce Eq.~(\ref{Four:1}) to the form
\begin{equation}\label{Four:2}
 \kappa\,\Delta T =  \frac{\partial T}{\partial t},
\end{equation}
where the thermal diffusivity $\kappa$, defined as
\begin{equation}\label{kap}
 \kappa = \frac{K}{\rho \, c_\mathrm{p}},
\end{equation}
will be assumed constant, leading to the homogeneous body thermal model.

The plane-parallel model (PPM) results from two simplifying assumptions: i) the penetration depth of the heat wave
is small compared with the radius of curvature for all fragments of the body surface, and
ii) there is no heat exchange in the direction perpendicular to the surface normal. Both the assumptions are plausible for
large objects with a low conductivity and a smooth, preferably convex surface.
In PPM we introduce the depth variable $\zeta$ whose values increase from  $\zeta= 0$ on the surface to higher positive values
inside the body. The basic equation of the homogenous body PPM is a reduced form of (\ref{Four:2})
\begin{equation}\label{Four:3}
 \kappa\,\frac{\partial^2 T}{\partial \zeta^2} =  \frac{\partial T}{\partial t},
\end{equation}
with nonlinear Robin boundary conditions
\begin{equation}\label{bc:3}
 \varepsilon_\mathrm{t} \, \sigma \, T^4(0) - K\, T'(0) - (1-A)\,E = 0,
\end{equation}
on the surface, and Neumann boundary condition in the limit of infinite depth
\begin{equation}\label{bc:4}
 \lim_{\zeta \rightarrow \infty}  T'(\zeta)  = 0.
\end{equation}
In both cases we use
\begin{equation}\label{dT}
   T' = \frac{\partial T}{\partial \zeta},
\end{equation}
and Eq.~(\ref{bc:3}) results from the energy balance (\ref{bc:1}) with
\begin{equation}\label{bc:10}
  \vv{\hat{n}} \cdot \nabla T = - T'.
\end{equation}
Accordingly, Eq.~(\ref{Tprim}) can be replaced by
\begin{equation}\label{Tprim:pp}
 Q_j =   K \, T'_j.
\end{equation}

Instead of initial conditions at some specified epoch $t$, we impose the quasi-periodicity
condition, requiring that all transient terms have been damped after sufficient relaxation time.
This condition is most easily imposed by assuming from the beginning that $T$ is replaced by its
DFT with respect to rotation phase $\Omega$ and mean anomaly $\ell$.

Since the assumptions of PPM exclude the heat transfer between adjacent triangles (and their associated volumes),
we may consider each body fragment separately, so the index referring to a particular face (like $j$ in Eq.~(\ref{Tprim:pp}))
will be omitted in the following discussion.

\subsection{Helmholtz equation and its solution}

\subsubsection{General case}

Let us consider the DFT of temperature
\begin{equation}\label{T:DFT}
    \vv{\hat{T}} = N^{-2}\,\mtr{F}_2 \vv{T}.
\end{equation}
Resorting to the associated trigonometric polynomial (\ref{trig}) substituted into Fourier equation (\ref{Four:3}),
we find that the DFT coefficients, as functions of  depth $\zeta$, obey a system of decoupled 1D Helmholtz equations
\begin{equation}\label{Helm:1}
 \left( \vv{\hat{T}}[p] \right)'' - \mathrm{i}\,\beta_p\, \vv{\hat{T}}[p] = 0, \qquad p = 0,\ldots,N^2-1,
\end{equation}
where $N$ is the angles sampling density, and parameters $\beta_p$ depend on orbital mean motion $\nu$ and rotation rate $\omega$;
if $p = j+k N$, then
\begin{equation}\label{beta}
    \beta_p = \frac{Z_N(j)\,\nu + Z_N(k)\,\omega}{\kappa},
\end{equation}
where $Z_N$ is defined in  Appendix~\ref{ap:A}.

Using a formal analogy with harmonic oscillator (with a complex frequency), and imposing the Neumann condition (\ref{bc:4})
we obtain a solution, depending on one arbitrary constant $C_p$, in a form
\begin{equation}\label{T:sol}
 \vv{\hat{T}}[p] = C_p\, \exp\left[ - \left(1+\mathrm{sgn}(\beta_p)\,\mathrm{i}  \right)\,\sqrt{\frac{|\beta_p|}{2}}\,\zeta \right].
\end{equation}
In principle, $C_p$ should now be determined from the second boundary condition, but for the further treatment we need
only the logarithmic derivative
\begin{equation}\label{T:rat}
    \gamma_p =  \frac{\vv{\hat{T}'}[p]}{\vv{\hat{T}}[p]} = - \left(1+  \mathrm{i}\,\mathrm{sgn}(\beta_p) \right)\,\sqrt{\frac{|\beta_p|}{2}},
\end{equation}
which occurs to be independent on $\zeta$ and allows to express the derivative $T'$  in terms of $T$.

\subsubsection{Null frequency}

\label{Nullf}
The general solution (\ref{T:sol}) is not valid for $\beta_p = 0$, when Eq.~(\ref{Helm:1}) degenerates into
\begin{equation}\label{Helm:0}
 \left( \vv{\hat{T}}[p] \right)''   = 0.
\end{equation}
It happens when $p=0$, i.e. for the mean value of temperature $T$.

The solution of (\ref{Helm:0}) is a linear function of $\zeta$, but matching it with the boundary condition
(\ref{bc:4}) we find that the null frequency solution is $\vv{\hat{T}}[0] = \mathrm{const}$, hence
\begin{equation}\label{gam:0}
    \gamma_0 = 0.
\end{equation}
The fact that $\gamma_0 = 0$,  has significant implications for the YORP influence on $\omega$.

\subsection{Boundary conditions}

\subsubsection{Newton-Raphson setup}

Knowing the ratios $\gamma_p$, we can find the spectrum  $\vv{\hat{Q}}$ from
the boundary conditions (\ref{bc:3}).
Consider the vector of sampled temperature values  $\vv{T}$ at the centroid of a given triangular face.
We will designate by $\vv{T}^{n}$ the vector of the $n$-th powers of $T$, i.e.
\begin{equation}\label{T4}
    \vv{T}^n[p] = \left( \vv{T}[p]\right)^n, \qquad p = 0, \ldots, N^2-1.
\end{equation}
Then, using the DFT formalism from Appendix~\ref{ap:A}, the boundary conditions (\ref{bc:3}) can be written in the vector form
\begin{equation}\label{bcv:1}
    \epsilon_\mathrm{t} \, \sigma \, \vv{T}^4  -  \vv{Q} = (1-A)\, \vv{E}.
\end{equation}
Using Eqs.~(\ref{DFTI}), (\ref{DFT}), (\ref{Tprim:pp}) and (\ref{T:rat}), we find that
\begin{equation}\label{bcv:2}
      \vv{T}^4  + N^{-2} \left( \mtr{F}^\ast_2 \mtr{B}\,\mtr{F}_2 \right) \vv{T} = \frac{1-A}{ \epsilon_\mathrm{t} \, \sigma}\,\vv{E},
\end{equation}
where $\mtr{B}$ is an $N^2 \times N^2$ diagonal matrix with
\begin{equation}\label{Bdef}
    B_{pp} =  - \frac{K\,\gamma_p}{\epsilon_\mathrm{t} \, \sigma}.
\end{equation}

The main difficulty in dealing with the energy balance equation (\ref{bcv:2})
is its nonlinearity, requiring the use of some approximate methods. Resorting to the Newton-Raphson
method, we can establish an iterative scheme
\begin{equation}\label{bcv:n}
     \left( \mtr{D}^{(m)} + \mtr{C} \right) \vv{T}^{(m+1)} = \vv{G}^{(m)},
\end{equation}
where $\mtr{D}$ is a diagonal matrix with
\begin{equation}\label{Ddef}
    D^{(m)}_{pp} = 4\,\left(\vv{T}^3[p]\right)^{(m)},
\end{equation}
$\mtr{C}$ is a normal, block circulant matrix
\begin{equation} \label{Cdef}
\mtr{C} =  N^{-2}\,\mtr{F}^\ast_2 \mtr{B}\,\mtr{F}_2,
\end{equation}
with all eigenvalues $\lambda_p = B_{pp}$ having non-negative real parts, and
\begin{equation}\label{Gdef}
 \vv{G}^{(m)} = \frac{1-A}{ \epsilon_\mathrm{t} \, \sigma}\,\vv{E}  + \frac{3}{4} \mtr{D}^{(m)}\,\vv{T}^{(m)}.
\end{equation}
In principle, starting from any reasonable approximation $\vv{T}^{(0)}$,
we can solve the linear system (\ref{bcv:n}), approaching a sufficiently accurate $\vv{T}$ with a quadratic convergence.
Then the spectrum $\vv{\hat{Q}}$, required in (\ref{M3:a1},\ref{M2:a1}), follows from
\begin{equation}\label{Q:DFT}
    \vv{\hat{Q}} = - \epsilon_\mathrm{t} \, \sigma\, N^{-2}\,\mtr{B}\,\mtr{F}_2 \vv{T} = - \epsilon_\mathrm{t} \, \sigma\, \mtr{B}\,\vv{\hat{T}},
\end{equation}
efficiently executed by one call of the fast Fourier transform (FFT) routine.

Regretfully, the left-hand side of Eq.~(\ref{bcv:n}) contains a dense,  $N^2 \times N^2$ matrix that cannot be directly
inverted by low cost algorithms. This is quite frustrating, because the inversion of the diagonal matrix
$\mtr{D}$ is trivial, while inverting $\mtr{F}^\ast_2 \mtr{B}\,\mtr{F}_2$ alone is easily done by the FFT.
But before we show the way to solve this problem, an important property of the nonlinear system (\ref{bcv:2})
is worth stating.

\subsubsection{Conductivity has no influence on the rotation period in the PPM}

\label{chne}
According to Sect.~\ref{Nullf} and Eq.~(\ref{Bdef}), the element $B_{00} = 0$. As a consequence, the first row of
the matrix $\mtr{F}^\ast_2 \mtr{B}$ contains only zeros, hence
\begin{equation}\label{Q0}
    \vv{\hat{Q}}[0] = 0.
\end{equation}
Thus, returning to Eq.~(\ref{M3:a1}) we conclude that, as far as the plane-parallel model of a homogeneous body is concerned,
the conductivity complement has no effect on the mean value of the YORP torque component responsible for $\dot{\omega}$, i.e.
\begin{equation}\label{M3:a0}
  \left\langle  \vv{M}^j \cdot \vv{e}_3 \right\rangle = -\frac{2}{3\,c} \,\vv{\hat{E}}_j[0] \,\left(\vv{r}_j \times \vv{S}_j \right) \cdot \vv{e}_z,
\end{equation}
is determined by the Rubincam part alone.

A similar observation was reported in previous works, although each time with different assumptions. \citet{Mysen:08}, \citet{NV:08}, and \citet{BM:08}
found it assuming the infinite radius of a homogeneous object, but they made additional assumptions of linearized temperature
variations and pseudo-convex shadowing model. Numerical results of \citet{CV:04}, using the assumptions similar to these of the present paper,
were nonconclusive: some objects seemed to have $\dot{\omega}$ independent on conductivity, but some (like 6489 Golevka) were exceptions from this rule.\footnote{\citet{CV:04} write about `a near independence' on $K$.}
The arguments based on the spectrum of derivative $\zeta'$, support our earlier conjecture \citep{BBCOV:09} that the apparent dependence on $K$
is definitely due to numerical errors -- most probably a too short relaxation time and/or inaccurate discretisation in the time stepping finite difference
scheme of \citet{CV:04}.

The first significant dependence of $\dot{\omega}$ on $K$ was announced in the analytical model of \citet{BVN:09} which allowed for a finite body radius
and used a 3D heat diffusion equation in spherical coordinates, although -- as usually in analytical models -- with many additional simplifications.
Thus, a central question is which of the two factors generates the dependence on conductivity. Appendix~\ref{Finde} presents the extension of
the PPM to the 1D model with finite radius; even in this generalization $\gamma_0 = 0$, hence we can state, that the necessary condition for the dependence
of the YORP effect in spin rate on conductivity is the heat exchange perpendicular to the surface normal, i.e. a 3D heat diffusion model.

\subsubsection{HN iterations}
\label{HNs}

In order to solve Eq.~(\ref{bcv:n}), we took the approach based upon the idea of \citet{HoNg:05} who considered circulant-plus-diagonal systems
with imaginary diagonal part $(\mathrm{i} \mtr{D} + \mtr{C})$. Unfortunately, major part of the proofs given by
Ho and Ng relies on the skew-Hermitian property of $\mathrm{i} \mtr{D}$, so we adopted their method to our
$(\mtr{D}+\mtr{C})$ system  \textit{faute de mieux}, hoping that HN iterations\footnote{An acronym equally matching the authors' names and
the Hermitian-plus-normal nature of the system.} will work anyway.

According to the HN algorithm, at each Newton step (\ref{bcv:n}) of the `outer iterations', one should introduce `inner iterations'
\begin{eqnarray}
\label{HN:1}
  \left(\tau\, \mtr{I} + \mtr{C}\right)\,\vv{Y}^{(k)} & = & \left(\tau \,\mtr{I} - \mtr{D}^{(m)}\right)\,\vv{T}^{(m,k)} + \vv{G}^{(m)},  \\
\label{HN:2}
  \left(\tau\, \mtr{I} + \mtr{D}^{(m)}\right)\,\vv{T}^{(m+1,k)} & = & \left(\tau \, \mtr{I} - \mtr{C}\right)\,\vv{Y}^{(k)} + \vv{G}^{(m)},
\end{eqnarray}
where $\tau > 0 $ is some arbitrary real parameter, and $\vv{Y} \in \mathbb{R}^{N^2}$ is an auxiliary vector.

Concatenating Eqs.~(\ref{HN:1}) and (\ref{HN:2}), one can see that the convergence of this process depends on the spectral
radius $\rho(\mtr{M})$ of the matrix
\begin{equation}\label{M:matr}
    \mtr{M} = \left(\tau\, \mtr{I} + \mtr{D} \right)^{-1}\,\left(\tau \, \mtr{I} - \mtr{C}\right)\,
    \left(\tau\, \mtr{I} + \mtr{C}\right)^{-1}\,\left(\tau \,\mtr{I} - \mtr{D} \right),
\end{equation}
(superscripts $(m)$ omitted) which is bounded by
\begin{equation}\label{spr}
    \rho(\mtr{M}) \leqslant \max_p \frac{|\tau-D_{pp}|}{|\tau+D_{pp}|}\,\max_p \frac{|\tau-B_{pp}|}{|\tau+B_{pp}|}.
\end{equation}
Knowing that either $B_{pp} = 0$, or $\Re{(B_{pp})} = |\Im{(B_{pp})}| > 0$, we conclude
\begin{equation}\label{mp}
    \max_p \frac{|\tau-B_{pp}|}{|\tau+B_{pp}|} = 1,
\end{equation}
hence
\begin{equation}\label{spr:2}
    \rho(\mtr{M}) \leqslant \max_p \frac{|\tau-D_{pp}|}{|\tau+D_{pp}|}.
\end{equation}
Thus the spectral radius is less than 1, provided the diagonal of $\mtr{D}$ does not contain zero or negative values of $4\,T^3$.
It means that HN iterations will converge faster at higher conductivity values, when the minimum temperature does not
drop significantly during the night. On the other hand, Eq.~(\ref{spr:2}) suggests a safe and nearly optimal
choice of $\tau$ as the geometric mean of the maximum and minimum diagonal entries of $\mtr{D}$
\begin{equation}\label{tau}
    \tau = \sqrt{D_{\max}\,D_{\min}}.
\end{equation}
This rule obviously fails for $D_{\min} = 0$. But, what is worse, the shadowing effects lead to discontinuities in the insolation, causing the
so called ringing artifacts -- often with negative values of temperature. From the point of view of upper bounds (\ref{spr:2}), the iterations should diverge
in such cases, but the algorithm occurs to be unexpectedly  robust and often converges in spite of $T <0$, although once the temperature drops below 0, a number of wild and
chaotic jumps can be observed before the residues resume their decreasing path. After a number of trials we have finally adopted
a practical rule of thumb
\begin{equation}\label{tau:1}
    \tau  = \max{\left(\sqrt{D_{\max}\,|D_{\min}|}, \, \sqrt{D_{\max}} \,\right)},
\end{equation}
handling the negative $D_{\min}$ case, and protecting $\tau$ from taking excessively small values.

\subsubsection{Quasi-Newton  method}

When using combined inner-outer iterations schemes, one always faces a problem when to terminate the inner loop before the improvements
become nonsignificant from the point of view of the current outer iteration. At this point we trade efficiency for simplicity
and retain only one inner HN step, obtaining the final quasi-Newton scheme with two substeps
\begin{eqnarray}
\label{FHN:1}
  \left(\tau^{(m)}\, \mtr{I} + \mtr{C}\right)\,\vv{Y}  & = & \left(\tau^{(m)} \,\mtr{I} - \mtr{D}^{(m)}\right)\,\vv{T}^{(m)} + \vv{G}^{(m)},  \\
\label{FHN:2}
  \left(\tau^{(m)}\, \mtr{I} + \mtr{D}^{(m)}\right)\,\vv{T}^{(m+1)} & = & \left(\tau^{(m)} \, \mtr{I} - \mtr{C}\right)\,\vv{Y}  + \vv{G}^{(m)},
\end{eqnarray}
where $\tau^{(m)}$, $\mtr{D}^{(m)}$, and $\vv{G}^{(m)}$ are recomputed at each step $m$ (but not between
the substeps (\ref{FHN:1}) and (\ref{FHN:2})).

In practical terms, solving the equations of the quasi-Newton method is quite simple and requires the storage of only few
1D arrays with $N^2$ elements. The matrix-vector product in the right-hand side of Eq.~(\ref{FHN:1}) is obviously executed in a single
$N^2$ loop, generating some vector $\vv{X} \in \mathbb{R}^{N^2}$ according to
\begin{equation}\label{X}
    \vv{X}[p] = \left( \tau^{(m)}- D^{(m)}_{pp} \right) \,\vv{T}^{(m)}[p] + \vv{G}^{(m)}[p],
\end{equation}
where $p=0,\ldots,N^2-1$.
We compute this vector and find its DFT $\vv{\hat{X}}$ according to the definition (\ref{DFT}).
In order to solve the system $\left(\tau^{(m)}\, \mtr{I} + \mtr{C}\right)\,\vv{Y} = \vv{X}$,
we note that
\begin{equation}\label{tau:I:C}
    \left(\tau^{(m)}\, \mtr{I} + \mtr{C}\right) \vv{Y} =
     \left(\mtr{F}^\ast_2\left(N^{-2}\,\tau^{(m)}\, \mtr{I}\right)\mtr{F}_2 + \mtr{C}\right) \vv{Y},
\end{equation}
so, substituting Eqs.~(\ref{Cdef}, \ref{X}, \ref{tau:I:C}) into (\ref{FHN:1}) we obtain
\begin{equation}\label{FHN:1a}
    \mtr{F}^\ast_2 \left(\tau^{(m)}\, \mtr{I} +  \mtr{B} \right)\,\vv{\hat{Y}} = \vv{X},
\end{equation}
where $\vv{\hat{Y}} = N^{-2} \mtr{F}_2 \vv{Y}$ is the DFT of $\vv{Y}$.
Thus, the first substep is completed by defining, but not yet evaluating, the transform $\vv{\hat{Y}}$
\begin{equation}\label{FHN:1b}
   \vv{\hat{Y}}[p] =   \frac{\vv{\hat{X}}[p]}{\tau^{(m)} +  B_{pp}}.
\end{equation}
Solving Eq.~(\ref{FHN:2}) we use a similar approach: first, the product in the right-hand side is expressed as
\begin{equation}\label{FHN:2a}
    \left(\tau^{(m)} \, \mtr{I} - \mtr{C}\right)\,\vv{Y}  = \mtr{F}^\ast_2 \left(
    \tau^{(m)}\, \mtr{I} -  \mtr{B}\right)\,\vv{\hat{Y}}.
\end{equation}
It means, that we have to compute the inverse DFT $\vv{W} = \mtr{F}^\ast_2 \vv{\hat{W}}$, where
\begin{equation}\label{W:def}
    \vv{\hat{W}}[p] = \frac{ \tau^{(m)} - B_{pp} }{\tau^{(m)} +  B_{pp}}\,\vv{\hat{X}}[p],
\end{equation}
and then we obtain the $m$-th approximation of surface temperature
\begin{equation}\label{Tm}
 \vv{T}^{(m)}[p] = \frac{\vv{W}[p] + \vv{G}^{(m)}[p]}{\tau^{(m)} + D^{(m)}_{pp}}.
\end{equation}

Each step of this process requires two calls of the FFT procedures, one direct ($\vv{X} \rightarrow  \vv{\hat{X}}$)
and one inverse ($\vv{\hat{W}} \rightarrow  \vv{W}$),
as well as few loops with $N^2$  complex products, which is probably not far from the optimum computational cost.

\subsection{First guess and accuracy}

The fundamental question accompanying each iteration process is how to start and when to stop.
It looks reasonable to assume the starting value  $\vv{T}^{(0)}$ by setting $K=0$ in the original, nonlinear
boundary conditions (\ref{bcv:2}), which leads the choice between
\begin{equation} \label{start:0}
\vv{T}^{(0)} =  \left(\frac{1-A}{ \epsilon_\mathrm{t} \sigma} \right)^\frac{1}{4}\,\vv{E}^\frac{1}{4},
\end{equation}
or, apparently simplistic,
\begin{equation} \label{start:c}
\vv{T}^{(0)}[p] =  \left( (1-A)\,\frac{\vv{\hat{E}}[0]}{\epsilon_\mathrm{t} \sigma}\right)^\frac{1}{4} = T_0, \qquad p = 0, \ldots, N^2-1.
\end{equation}
Choosing a constant $\vv{T}^{(0)}$ according to (\ref{start:c}) may seem too crude, since it means that iterations will have to reconstruct
all periodic terms with leading amplitudes -- in the worst case of low conductivity -- comparable in magnitude to the mean value.
But the practice shows a superiority of (\ref{start:c}) over (\ref{start:0}). Building the amplitudes up from zero is numerically more stable
than decreasing their values from the state, when the temperature determined by (\ref{start:0}) takes zero values. This fact can be explained from a number of
points of view, using both physical and mathematical arguments. Focusing on the latter, note that
according to the estimates given in Sect.~\ref{HNs}, the spectral radius $\rho(\mtr{M})$ equals 1 when any of $\vv{T}^{(0)}[p]=0$.
Moreover, the approximation (\ref{start:0}) is a continuous, but not smooth function of $\ell$ and $\Omega$, which significantly
degrades the numerical quality of the DFT of its derivative with respect to these angles.

Let us write explicitly the values of $\vv{\hat{T}}^{(1)}$ resulting from the quasi-Newton iterations when $\vv{T}^{(0)}$ is given by Eq.~(\ref{start:c}).
In this case, $\vv{\hat{T}}^{(0)}[p] = 0$ for all $p \neq 0$, the mean value is $\vv{\hat{T}}^{(0)}[0] = T_0$,  and diagonal matrix $\mtr{D}^{(0)} = 4\,T_0^3\,\mtr{I}$,
hence $\tau^{(0)} = 4\,T_0^3$. Using
\begin{equation}\label{G0}
 \vv{G}^{(0)} = \frac{1-A}{ \epsilon_\mathrm{t} \, \sigma}\,\vv{E}  + 3\,T_0^3\,\vv{T}^{(0)},
\end{equation}
we obtain from (\ref{FHN:1}) and (\ref{FHN:2}), left multiplied by $\mtr{F}_2$,
\begin{eqnarray}
   \vv{\hat{T}}^{(1)}[0] & = & N^{-2}\,T_0, \label{T1:0}\\
   \vv{\hat{T}}^{(1)}[p] & = &  \frac{(1-A)}{\epsilon_\mathrm{t} \sigma}\,\frac{\vv{\hat{E}}[p]}{4\,T_0^3 + B_{pp}}, \qquad p = 1, \ldots, N^2-1.
\label{T1:p}
\end{eqnarray}
Remarkably, the same result can be obtained even easier from the original Newton-Raphson system (\ref{bcv:n}).
Equations (\ref{T1:0}) and (\ref{T1:p}) define the linear thermal model -- a standard tool in analytical YORP theories.
Of course, the direct application of Eqs.~(\ref{T1:0}) and (\ref{T1:p}), followed by one FFT call to obtain $\vv{T}^{(1)}$ is much cheaper
than performing the first iteration of (\ref{FHN:1}) and (\ref{FHN:2}) in its regular form. For this reason we actually start iterations
from $m=1$, obtaining the means to simulate the results of linear model as an extra profit.

The iterations cycle has to be stopped when a sufficient accuracy is attained. The stopping criterion should be chosen carefully. The simplest one is to observe
the differences between subsequent values of $\vv{\hat{T}}^{(m)}[1]$ and exit when $|\vv{\hat{T}}^{(m)}[1]- \vv{\hat{T}}^{(m-1)}[1]| < \delta $. But one has to be careful, because
when the convergence is slow (a typical situation at low conductivities), such a difference carries no information about the accuracy of $\vv{\hat{T}}^{(m)}[1]$.
Fortunately, we have also an objective criterion $\epsilon_\mathrm{t} \sigma \left\langle \vv{T}^4 \right\rangle = (1-A)\,\vv{\hat{E}}[0]$, independent on the convergence rate.
And since the mean value of $\vv{T}^4$ accumulates also the errors of all periodic terms of $\vv{T}$, we exit the iterations when, for a specified temperature
tolerance $\delta$, two conditions are simultaneously satisfied:
\begin{equation}\label{cnd:exit}
 \left|\vv{\hat{T}}^{(m)}[1]- \vv{\hat{T}}^{(m-1)}[1] \right| \leqslant
 T_0^{-3}\,\left| \left\langle \left(\vv{T}^{(m)} \right)^4\right\rangle -  \frac{(1-A)}{\epsilon_\mathrm{t} \sigma}\,\vv{\hat{E}}[0] \right| < \delta.
\end{equation}

\section{Test runs}
\label{tests}

\subsection{Test bodies and accuracy requirements}

Two asteroids were chosen as test bodies for our numerical simulations: $1998~\mathrm{KY_{26}}$ with a relatively regular shape, and 6489 Golevka, whose large scale concavities
and sharp corners make it a good benchmark for the YORP models. Physical parameters adopted for the simulation are given in Tab.~\ref{tab:1}.
Radar shape models of both objects\footnote{Downloaded from the site  \texttt{http://www.psi.edu/pds/asteroid/} file version EAR\_A\_5\_DDR\_RADARSHAPE\_MODELS\_V2\_0.zip}
($4092$ triangular faces) were reduced to the center of mass and principal axes system assuming a constant density.

\begin{table}
\centering
\caption{Parameters of test objects}
\begin{tabular}{llcrr}
  \hline
   \multicolumn{3}{c}{} & $1998~\mathrm{KY_{26}}$ & 6489 Golevka \\
   \hline
  albedo & $A$ &  & $0.1$ & $0.1$ \\
  emissivity & $\epsilon_\mathrm{t}$ &  & $0.9$ & $0.9$ \\
  density & $\rho$ & $\mathrm{kg\,m^{-3}}$ & $2800$ & $2700$ \\
 specific heat & $c_\mathrm{p}$ & $\mathrm{J\,kg^{-1}\,K^{-1}}$ & $680$ & $680$ \\
 mom. inertia & $C$ & $\mathrm{kg\,m^{2}}$ & $1.9346 \times 10^9$ & $7.3420 \times 10^{15}$ \\
 rotation period & $P$ & $\mathrm{h}$ & $0.1748$ & $6.0264$ \\
 orbit semi-axis & $a$ & $\mathrm{au}$ & $1.232$ & $2.5$ \\
 eccentricity & $e$ &  & $0.2$ & $0.6$ \\
 approx. diameter &  & $\mathrm{m}$ & $\sim 30$ &$\sim 530$ \\
  \hline
\end{tabular}
\label{tab:1}
\end{table}

In most of previous YORP models either the orbits were simply assumed circular, or the YORP effect computed on a circular orbit was multiplied by
a conversion factor
\begin{equation}\label{conv}
    q_\mathrm{e} = \left(1 - e^2 \right)^{-\frac{1}{2}}.
\end{equation}
From theoretical standpoint, the latter procedure can be justified exact in the Rubincam's approximation or in linear thermal models, where
rotation and orbital motion effects are separable, but there are no reasons to assert it in general. The factor $q_\mathrm{e}$ concerns
all terms proportional to the average power flux (i.e. the ones with $\langle T^4 \rangle $) but not those depending on the first power of temperature
\citep{Rub:2004}.

In all computations we have adopted a rule that no error bars should be required in YORP plots. Various levels of sampling $N$ and tolerance $\delta$ had been tried
until a difference from the results with sampling $2N$ and tolerance $\delta/10$ became comparable to the plot line thickness. Finally, for $1998~\mathrm{KY}_{26}$
we used $N=256$, $\delta = 10^{-4}$, whereas Golevka, as expected, was more challenging and required $N=512$, $\delta = 10^{-4}$, or even $10^{-6}$, depending on conductivity.
For  the pseudo-convex shadowing model the requirements were less severe and $N$ could be two times smaller. Formal accuracy of the results presented in  next sections
is the following: $1998~\mathrm{KY}_{26}$ -- $3 \times 10^{-5}~\mathrm{rad\,s^{-1}\,My^{-1}}$ for the rotation rate, and $3 \times 10^{-4}~\mathrm{rad\,s^{-1}\,My^{-1}}$ for the attitude YORP;
6489 Golevka -- $0.03~\mathrm{rad\,d^{-1}\,My^{-1}}$ for $\dot{\omega}$, and $0.5~\mathrm{rad\,d^{-1}\,My^{-1}}$ for the attitude effect.

\subsection{YORP effect in rotation rate}

\begin{figure}
 \begin{center}
  \includegraphics[width=8.5cm]{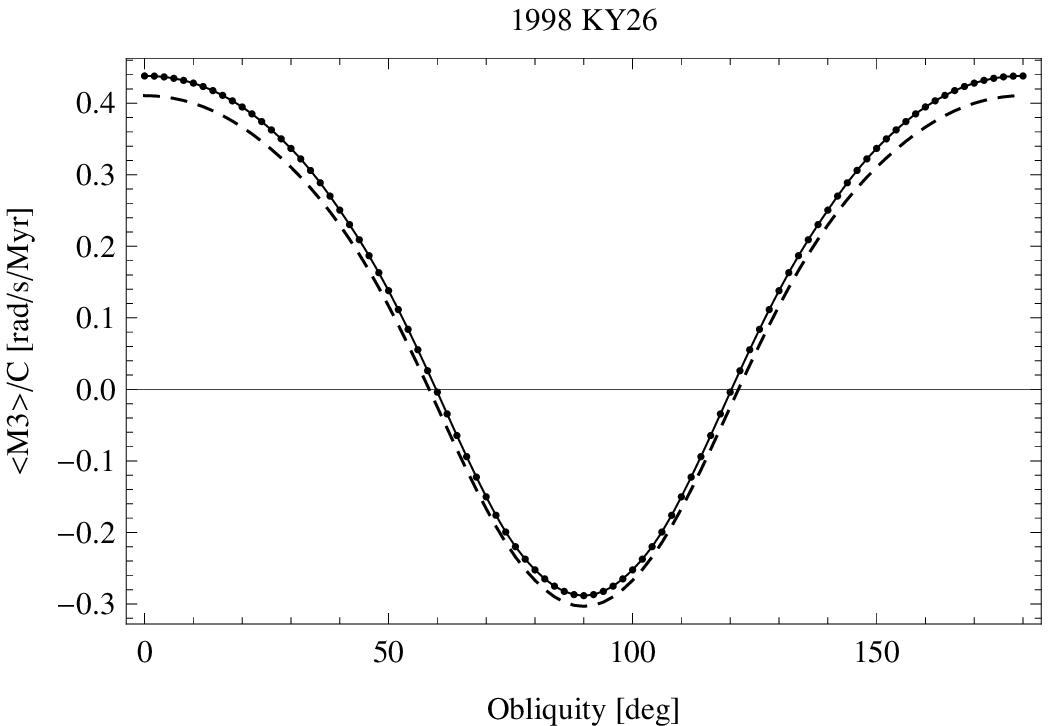}\\
  ~\\
  \includegraphics[width=8.5cm]{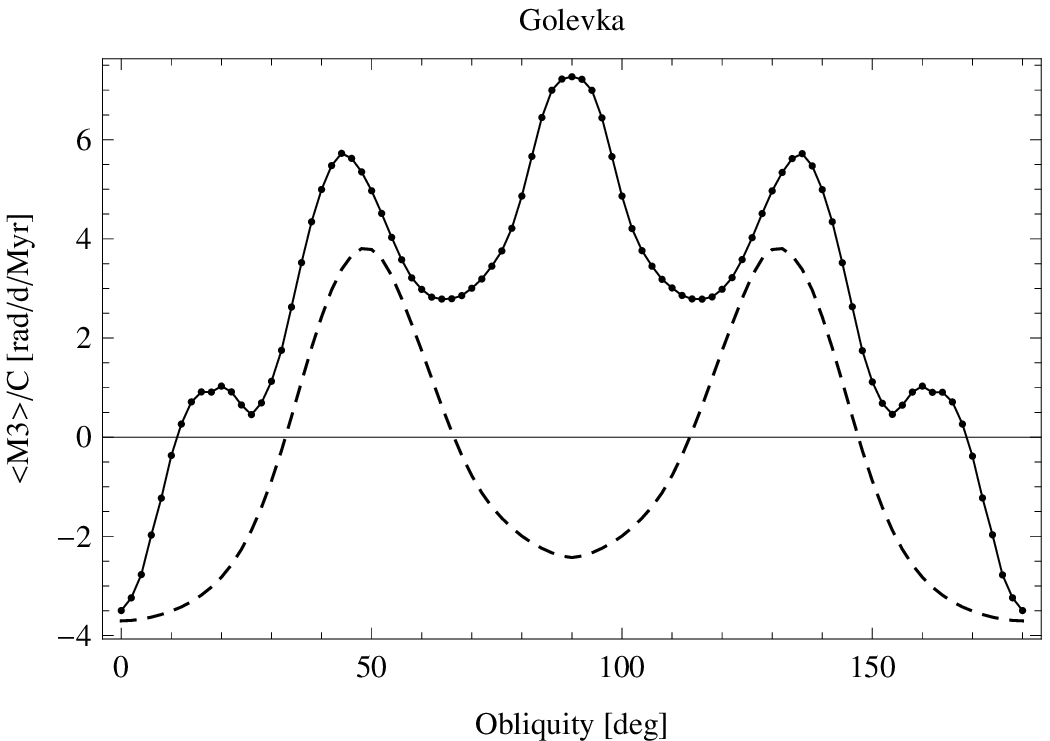}\\
  \end{center}
  \caption{YORP effect in rotation rate for $1998~\mathrm{KY_{26}}$ (top) and 6489 Golevka (bottom). Joined dots mark the values computed
  with $2\degr$ spacing in obliquity. Dashed line represents the pseudo-convex approximation. Note the difference in units between the top and bottom panels.}
  \label{fig:1}
\end{figure}

As we demonstrated in Sect.~\ref{chne} and Appendices~\ref{Regol} and \ref{Finde}, all kinds of 1D thermal models lead to the same YORP effect
in rotation rate, equivalent with the Rubincam approximation $K=0$. Figure~\ref{fig:1} shows the values of doubly averaged
$\left\langle M_3 \right\rangle C^{-1}$, where $M_3 = \vv{M} \cdot \vv{e}_3$.
According to Eq.~(\ref{eqm:1}), these values are equal to angular acceleration $\dot{\omega}$.
The dots in Fig.~\ref{fig:1} are placed for actually computed values of $\dot{\omega}$, and they form curves that fairly well agree with the results of \citet{VC:02},
provided the factor $q_\mathrm{e}$ is used and the differences in $C$ and $a_\mathrm{o}$ are accounted for.

The pseudo-convex approximation looks decent for a regular object like $1998~\mathrm{KY_{26}}$, but it fails completely for irregularly shaped Golevka.
It is worth noting, that the influence of shadowing for $1998~\mathrm{KY_{26}}$ amounts in principle to a vertical translation of the curve; a similar (although more prominent)
phenomenon was observed for 25143 Itokawa \citep{BBCOV:09}.

\subsection{YORP in attitude: seasonal effect revealed}

\begin{figure*}
 \begin{center}
  \includegraphics[width=17cm]{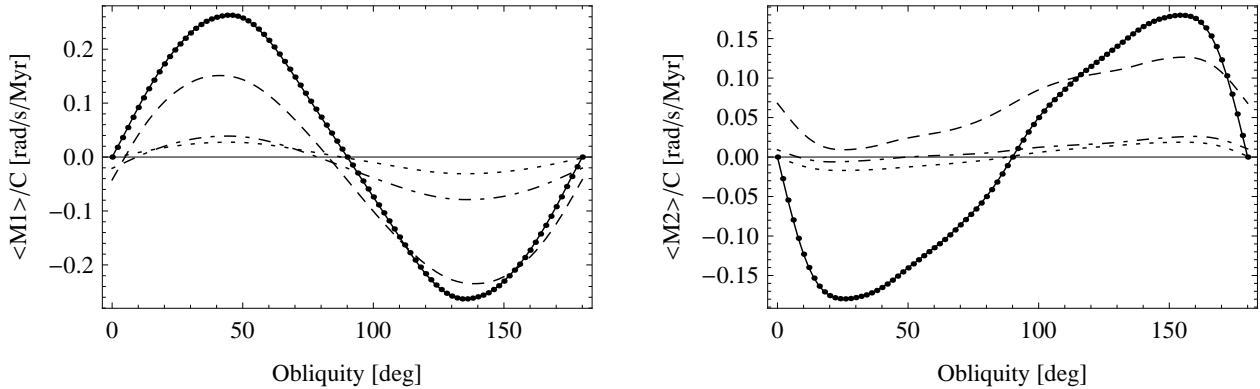}
  \end{center}
  \caption{YORP effect in obliquity (left) and precession (right) for $1998~\mathrm{KY_{26}}$ with $e=0.2$ and $\omega_\mathrm{o}=0$.
  Joined dots mark the actually computed values in Rubincam's approximation ($K=0$). Dashed, dash-dotted and dotted curves refer to the values
  from nonlinear 1D model at $K= 0.001$, $0.1$, and $10$ $\mathrm{W m^{1} K^{-1}}$, respectively. }
  \label{fig:2}
\end{figure*}

\begin{figure*}
 \begin{center}
  \includegraphics[width=17cm]{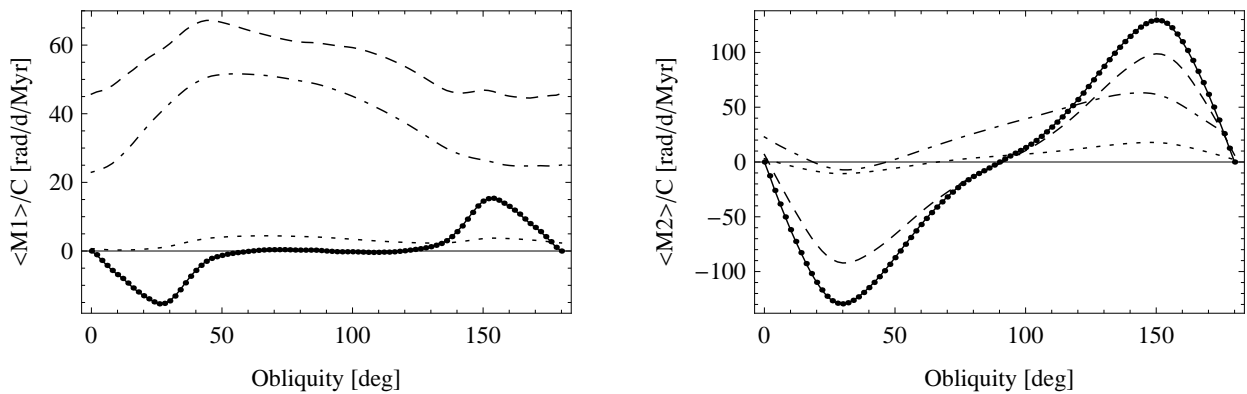}
  \end{center}
  \caption{Same as Fig.~\ref{fig:2} for 6489 Golevka with $e=0.6$ and $\omega_\mathrm{o}=0$. }
  \label{fig:3}
\end{figure*}

\begin{figure*}
 \begin{center}
  \includegraphics[width=17cm]{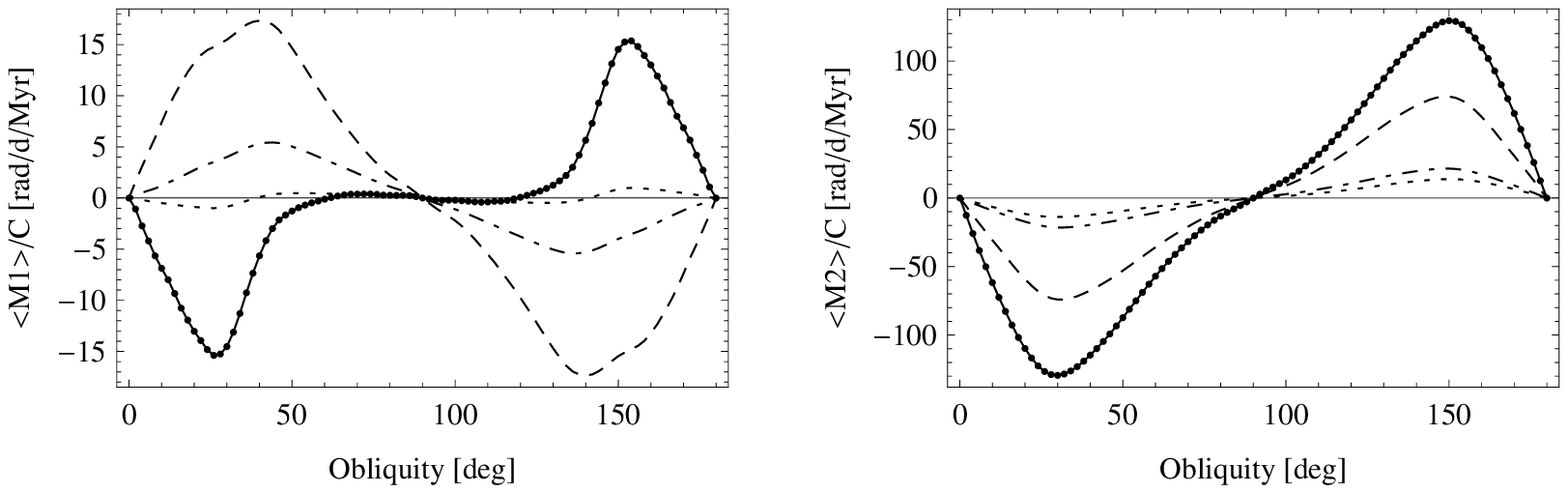}\\
  \includegraphics[width=17cm]{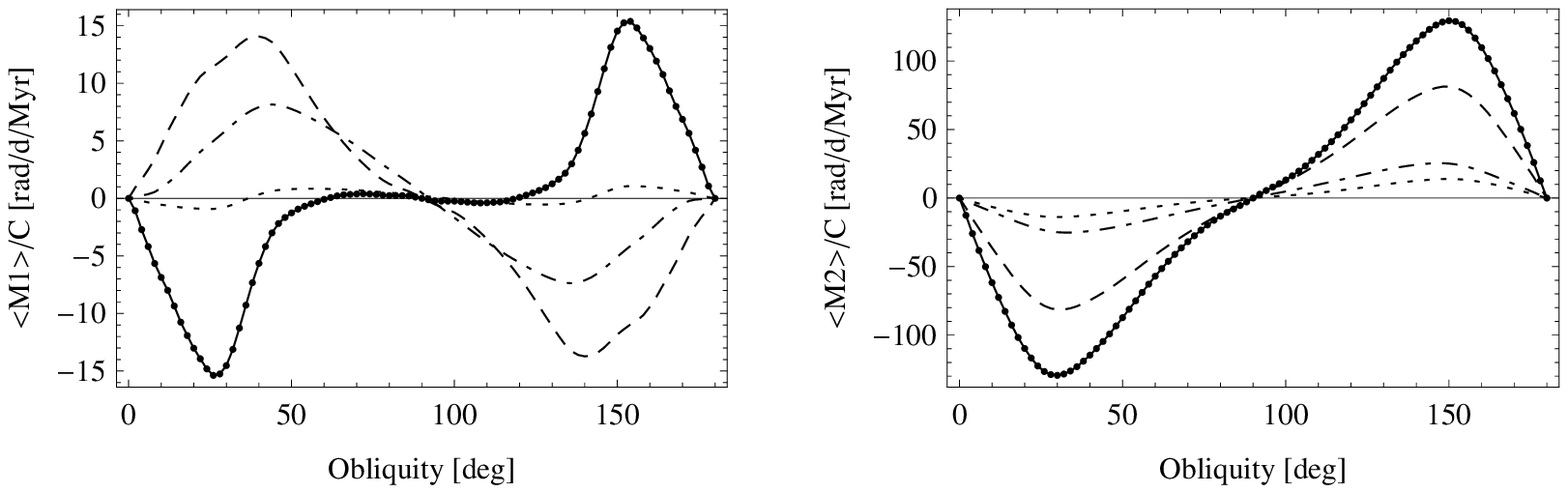}
  \end{center}
  \caption{YORP effect in obliquity (left) and precession (right) for 6489 Golevka. Top: linear approximation with $e=0.6$ and $\omega_\mathrm{o}=0$,
  bottom: circular orbit results rescaled by $\left(1-e^2\right)^{-\frac{1}{2}} = 1.25$. Conductivity like in previous figures.}
  \label{fig:4}
\end{figure*}

The YORP effect in attitude is usually described in terms of $\left\langle \vv{M}\cdot \vv{e}_1\right\rangle = \left\langle M_1 \right\rangle$,
and $\left\langle \vv{M}\cdot \vv{e}_2 \right\rangle = \left\langle \vv{M}_2 \right\rangle$. According to Sect.~\ref{dynam}, the mean value
of the drift in obliquity is given by $\omega\,\dot{\varepsilon} = \left\langle M_1 \right\rangle C^{-1}$, whereas $\left\langle M_2 \right\rangle C^{-1} =
\tan{\varepsilon} \left( \dot{\Omega} - \omega \right) $ is responsible for the mean precession component of the effect.
Figures~\ref{fig:2} and \ref{fig:3} demonstrate the influence of conductivity on these two attitude components within the nonlinear 1D thermal model.
Except for the Rubincam case, where a good agreement of $\left\langle M_1 \right\rangle$ with \citet{VC:02} is observed, the results may look surprising, not to say
ridiculous, at the first glance. Isn't it absurd to have $\dot{\varepsilon} > 0 $ for $\varepsilon = 180\degr$~? Why are the present curves so
different from all previously reported plots~?

The first question is relatively easy to resolve, since it is related to the classical problem of polar coordinates singularity close to the origin,
where a wrong parametrization may contradict physical facts. Nonzero mean values of $\left\langle M_1 \right\rangle$ and $\left\langle M_2 \right\rangle$
at $\sin{\varepsilon}=0$ merely indicate that the orientation of the spin axis normal to the orbital plane is not an equilibrium. A proper treatment of
passing through this state requires a formulation in terms of the spin vector and torque Cartesian coordinates (e.g.~\cite{BNV:05}).

As for the second question, we have to note that the previous theories of YORP with nonzero conductivity were mostly linear, approximating $T^4$ by $T_0^4 + 4 T_0^3\,T_1$,
with a constant $T_0$ and a purely periodic $T_1$. The only exception from this rule is the numerical model of \citet{CV:04}, but there the authors present only the results
for circular orbits. They did compute the values for $e\neq 0$ as well, but only for single, specific $\varepsilon$ and $\omega_\mathrm{o}$ pairs of actual objects and no
plots covering the whole range of obliquities have been published as yet.
Figure \ref{fig:4} shows that linear approximation generated by our model (top) and nonlinear results with $e=0$ (bottom)
behave exactly like in previous publications (except for a more complicated shape resulting from a better sampling than the 9 points interpolation of \citet{CV:04}).
Even a weak asymmetry of the obliquity YORP curve with respect to $\varepsilon = 90^\circ$ agrees with \citep{CV:04}.
These results imply that the shape of curves in Figs.~\ref{fig:2} and \ref{fig:3} is due to nonlinear coupling between daily and seasonal waves, and the effect
must be due to the variation of heliocentric distance, because the temperature variations due to change of seasons are present also in circular motion where nothing
unusual happens.

\begin{figure*}
 \begin{center}
  \includegraphics[width=17cm]{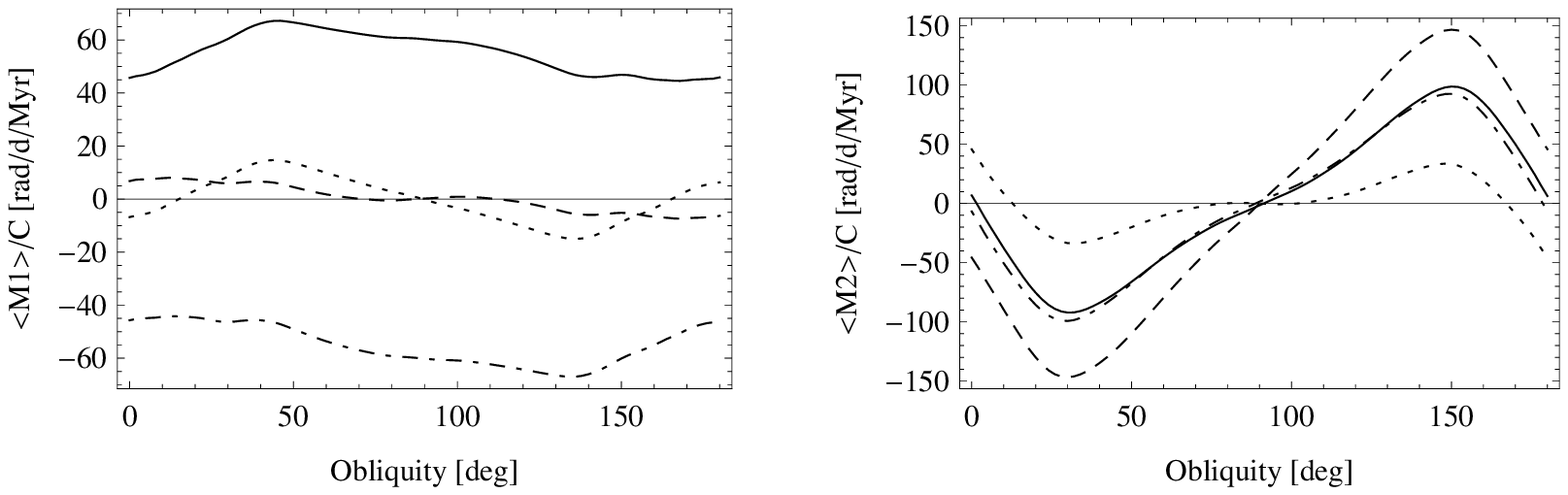}
  \end{center}
  \caption{YORP effect in obliquity (left) and precession (right) for 6489 Golevka on an eccentric orbit ($e=0.6$) with $\omega_\mathrm{o}=0$ (solid),
  $90\degr$ (dotted), $180\degr$ (dot-dashed), and $270\degr$ (dashed). Conductivity $K=10^{-3}~\mathrm{W\,m^{-1}\,K^{-1}}$.}
  \label{fig:5}
\end{figure*}

\begin{figure*}
 \begin{center}
  \includegraphics[width=17cm]{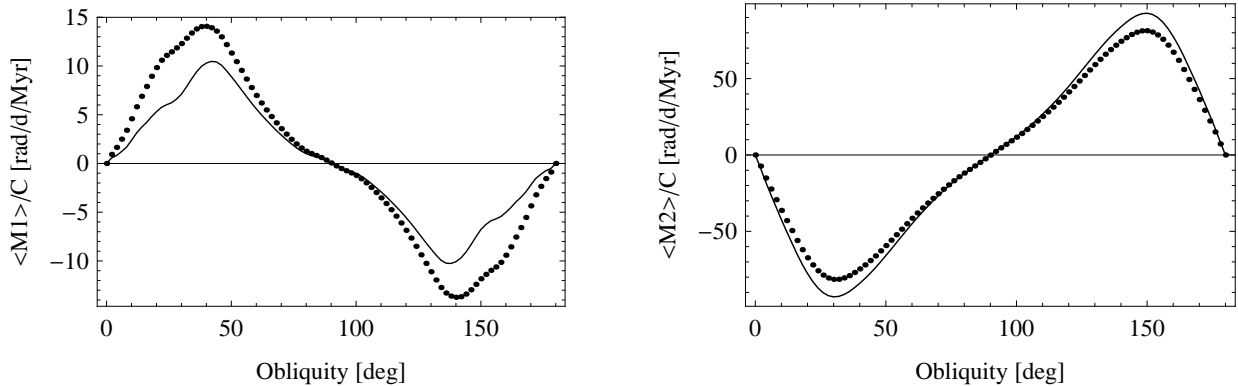}
  \end{center}
  \caption{Arithmetic mean of the four curves from Fig.~\ref{fig:5} (solid line) and the results for a circular orbit multiplied by $1.25$ (dots).}
  \label{fig:6}
\end{figure*}

\begin{figure*}
 \begin{center}
  \includegraphics[width=17cm]{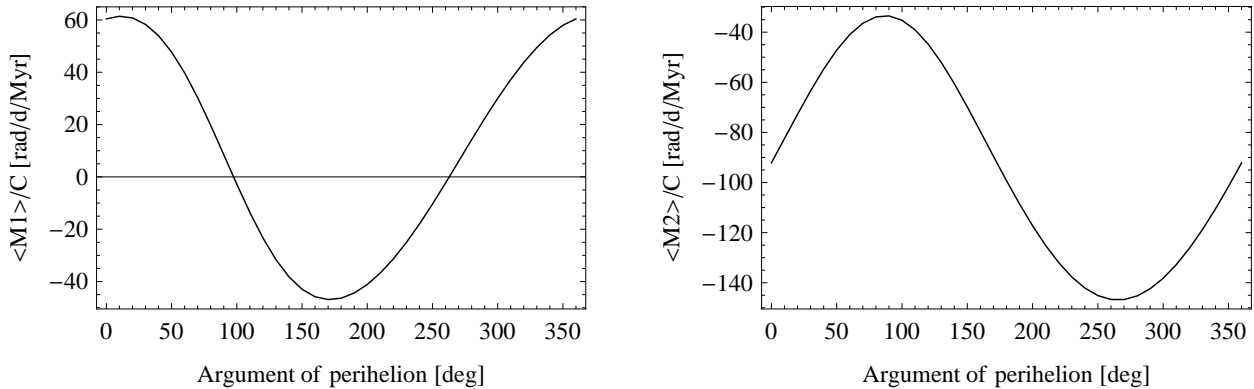}
  \end{center}
  \caption{YORP effect in obliquity (left) and precession (right) for 6489 Golevka on an eccentric orbit ($e=0.6$) with $\varepsilon = 30\degr$ and conductivity
  $K=10^{-3}~\mathrm{W\,m^{-1}\,K^{-1}}$ as a function of argument of perihelion $\omega_\mathrm{o}$.}
  \label{fig:7}
\end{figure*}

More light can be shed on the problem when changing the argument of perihelion $\omega_\mathrm{o}$, which was set to 0 in all previous plots.
Figure~\ref{fig:5} presents the attitude YORP effect for Golevka ($e=0.6$, $K=10^{-3}~\mathrm{W\,m^{-1}\,K^{-1}}$) with four different arguments
of perihelion $\omega_\mathrm{o}$ values ($0$, $90$, $180$, and $270$ degrees). Figure~\ref{fig:6} compares the arithmetic mean of the
four values with the results obtained for the circular orbit and re-scaled to $e=0.6$. On the other hand, Fig.~\ref{fig:7} shows the dependence
of the attitude YORP effect on the argument of perihelion (sampled by $4\degr$) when we fix the obliquity of Golevka at $\varepsilon = 30\degr$.
The dependence is almost (but not exactly) sinusoidal and the amplitude depends on eccentricity, although the dependence does not seem to obey a simple power law.

\subsection{The two layers model}

\begin{figure*}
 \begin{center}
  \includegraphics[width=17cm]{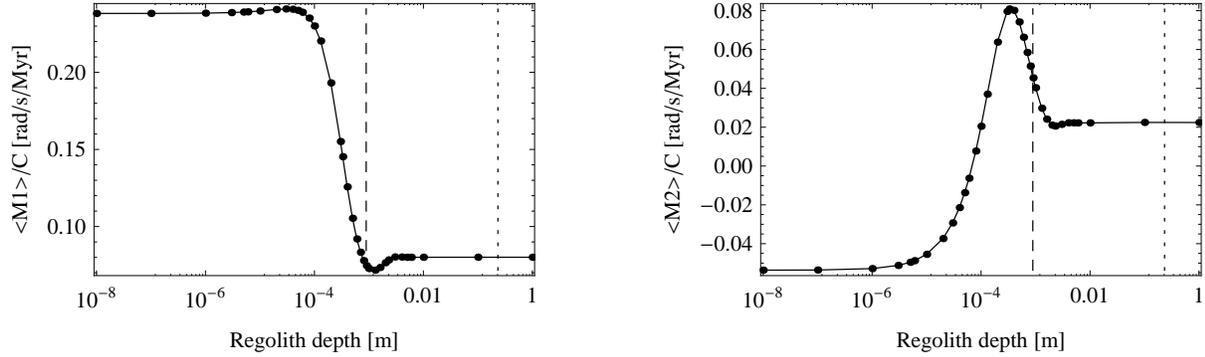}
  \end{center}
  \caption{YORP effect in obliquity (left) and precession (right) for $1998~\mathrm{KY_{26}}$ at $\varepsilon = 45\degr$ as a function of regolith depth $h$.
  Rotational and orbital thermal penetration depths are indicated by vertical lines (dashed and dotted, respectively).}
  \label{fig:r1}
\end{figure*}

\begin{figure*}
 \begin{center}
  \includegraphics[width=17cm]{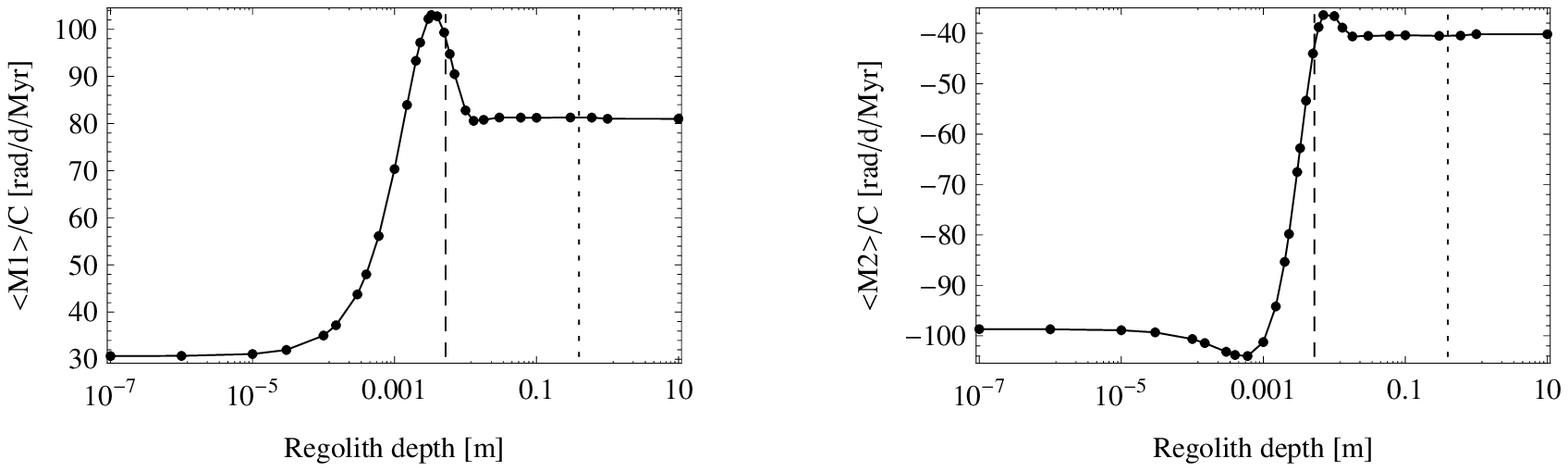}
  \end{center}
  \caption{Same as Fig.~\ref{fig:r1} for 6489 Golevka.}
  \label{fig:r2}
\end{figure*}

Introducing a more advanced model with a monolithic core and a regolith layer, described in Appendix~\ref{Regol}, has no influence on the YORP
effect in rotation rate. Thus, the results shown in Figs.~\ref{fig:r1} and \ref{fig:r2} concern only the effect in attitude.
In the test runs we have considered the values of density for both objects given in Tab.~\ref{tab:1} as the bulk densities serving to compute the moments of inertia,
but they are no longer used to compute the thermal diffusivity. Instead, we have adopted the following regolith parameters: $K = 0.01~\mathrm{W\,m^{-1}\,K^{-1}}$,
$c_\mathrm{P} = 760~\mathrm{J\,kg^{-1}\,K^{-1}}$, and $\rho = 1660~\mathrm{kg\,m^{-3}}$ \citep{RFC:08}. Accordingly, the thermal  diffusivity
of regolith layer is $\kappa \approx 7.93 \times 10^{-9}~\mathrm{m^2\,s^{-1}}$ and the physical properties of the core are specified by the ratio $w$
defined by Eq.~(\ref{nu}). We have assumed the value of $w = 0.1$ as a presumably realistic estimate.

Fixing a randomly chosen obliquity $\varepsilon = 45\degr$, we used our model to compute the YORP effect in attitude for various regolith depths $h$.
As it might be expected, there is a gradual transition between the thin and thick regolith cover results, and the curves in Figs.~\ref{fig:r1} and \ref{fig:r2} are
practically flat when prolonged towards higher or lower $h$ values. However, the transition is not monotonic, resembling a superposition of a logistic curve with
damped oscillations. This effect is understandable, observing that $h$ factors both the real and the imaginary parts of the exponential in
Eqs.~(\ref{T:rat:2},\ref{Hp}).  A similar pattern was present in the Yarkovsky force model of \citet[Fig.~A.8]{Cap:08} -- the only analogue that we can refer to.

The characteristic order of magnitude for the depth $h$ determining the transition from the thick to thin regolith case is the skin depth: a function of thermal diffusivity and
insolation frequency \citep{Lag:96}. However, there are two different principal skin depths  in our model: rotational $l_\mathrm{r}$, involving $\omega$, and orbital $l_\mathrm{o}$,
involving the mean motion $\nu$:
\begin{equation}\label{sd}
    l_{\mathrm{r}} = \sqrt{\frac{\kappa}{\omega}}, \qquad l_{\mathrm{o}} = \sqrt{\frac{\kappa}{\nu}}.
\end{equation}
Vertical lines in Figs.~\ref{fig:r1} and \ref{fig:r2} mark these two parameters (dashed for $l_{\mathrm{r}}$ and dotted for $l_{\mathrm{o}}$), indicating that
rotational skin depth (much smaller than $l_\mathrm{o}$) is the only important quantity. However, a significant deviation from the thin regolith mode occurs
already at the values of $h$ below $0.1\,l_\mathrm{r}$ or even $0.01\,l_\mathrm{r}$. With $l_\mathrm{r} \approx 1\,\mathrm{mm}$ for $1998~\mathrm{KY_{26}}$, and
$l_\mathrm{r} \approx 5\,\mathrm{mm}$ for Golevka, we can observe that the strongest dependence of the attitude YORP effect on $h$ is
observed when the regolith thickness is in the range of $0.1 \div 10 ~\mathrm{mm}$, which is quite similar to the results of \citet{VB:99} concerning
the Yarkovsky effect.

\section{Conclusions}

Thanks to the application of Fourier transform, the algorithm presented in this paper is more efficient and more accurate (although less general)
than its equivalent described by \citet{CV:04} and \citet{Cap:08}. Abandoning the finite difference approach in favor of
using exact solutions of the Helmholtz equation helped us to demonstrate, that the YORP effect in rotation period is the same in the Rubincam's
approximation ($K=0$) and in various 1D thermal models. As long as one neglects the heat transfer between adjacent surface elements,
the values of $\dot{\omega}$ do not depend on conductivity, regardless of  the body size and radial homogeneity assumptions.
From the point of view of observational detection of YORP, always based upon $\dot{\omega}$, this is a nice conclusion; not only because the
Rubincam's model is easier to compute, but also it requires less physical parameters to be known, since then -- at least for the Lambertian scattering and emission --
emissivity and albedo values do not matter. As  a matter of fact, the conclusion can be also given a straightforward physical explanation. If the heat conduction is restricted to the direction
normal to the surface, a nonzero mean value of the the temperature normal gradient $Q$ should imply systematic heating or cooling of asteroid's
interior. Hence, the property that $\langle Q \rangle = 0$ follows directly from the request of the energetic equilibrium state with transient terms relaxed.
However, according to the analytical model of \citet{BVN:09}, the YORP torque in rotation period for smaller bodies
with a 3D thermal model may differ from the Rubincam's approximation because of the heat flow between adjacent surface pathes that may receive a different mean power flux.

Even if the YORP effect occurred to be insensitive to the transverse heat conduction, the 1D models considered in this paper should not be applied to
objects whose diameter is small when compared with a skin depth. This restriction, explicitly stipulated in Sect.~\ref{sect:PPM}, can be physically explained
as follows. Consider a bar passing through the centre of a body  $O$ and intersecting the surface in two antipodal areas $S_1$ and $S_2$. The 1D models consider it
as two disjoint slabs with the absence of heat conduction at $O$ imposed as a boundary condition. In these circumstances, even if the conductivity
is very high, there is no possibility to transfer the heat from the sunlit $S_1$ to the dark $S_2$ in order to reach a smoother surface temperature distribution
and reduce the YORP strength. A possible improvement of 1D models might be based upon considering
the set of antipodal bars without the central cut; yet, in our opinion, a future investment in a complete 3D model is more needed.

The YORP effect in attitude is not directly observable, but still important for the simulations of long-term spin axis dynamics. The most prominent example is the analysis of
the Slivan states in Koronis family \citep{VNB:03}, considered the first, indirect proof of the YORP effect existence and significance.
Our results indicate that for elliptic orbits there exists a phenomenon that may be called a seasonal YORP effect in attitude by analogy with
the seasonal Yarkovsky effect in orbital motion \citep{Rub:95,VF:99}. The seasonal effect did not appear in earlier works based upon linearized thermal models, which led
to a hasty rule that the influence of orbital eccentricity amounts merely to a multiplicative factor from Eq.~(\ref{conv}). We confirm the validity of this rule
for the rotation period YORP, but not for the attitude. The effect passed unnoticed in the model of \citet{CV:04}, which can probably be explained by its high computational
time demands that discouraged experiments with various argument of perihelion values. The seasonal YORP in attitude deserves a closer inspection within a nonlinear analytical model
(even with a crude insolation model) that might help to explain its physical meaning. It is quite possible that there exists some relation between the seasonal YORP
and the Seversmith psychroterms mechanism discovered by \citet{Rub:2004}

As a final remark, let us observe that the presented model can be easily adapted to compute the Yarkovsky effect in orbital motion, like in the paper of \citet{Mysen:08}.

\section*{Acknowledgments}

The paper owes much to discussions with Dr. David Vokrouhlick\'{y}. The reviewer's comments by Dr. David Rubincam
calling for more physics behind the mathematical models are highly appreciated.
The work of S. Breiter was supported by the Polish Ministry of Science and Higher Education -- grant N N203 302535.

\bibliographystyle{mn2e}

\appendix

\section[]{Discrete Fourier transform}
\label{ap:A}

Our implementation of the algorithms presented in this paper relies on the extensive use of the
FFTW library (v. 3.2.2) developed by \citet{FFTW05}. The following formulae will use sign conventions,
normalization factors and 1D storage of 2D matrices (including the numbering of elements from 0)
adhering to the FFTW.

The size $N$ Fourier matrix $\mtr{F}$ is defined in terms of powers of
\begin{equation}\label{omfour}
    \omega_{j} = \mathrm{e}^{-\mathrm{i}\,j\,2\pi/N}, \quad j=0,\ldots,N-1,
\end{equation}
as
\begin{equation}\label{F1}
    \mtr{F} = \left(
     \begin{array}{ccc}
     \omega_0^0 & \ldots & \omega_0^{N-1} \\
     \vdots & \ddots & \vdots \\
     \omega_{N-1}^0 & \ldots & \omega_{N-1}^{N-1} \\
     \end{array}
      \right).
\end{equation}
For the 2D discrete Fourier transform (DFT) of the size $N \times N$, we define
the matrix $\mtr{F}_2$ using the Kronecker tensor product
\begin{equation}\label{F2}
    \mtr{F}_2 = \mtr{F} \otimes \mtr{F},
\end{equation}
with a resulting block structure of the $N^2 \times N^2$ matrix
\begin{equation}\label{F2b}
    \mtr{F}_2 =  \left(
     \begin{array}{ccc}
     \omega_0^0 \mtr{F} & \ldots & \omega_0^{N-1} \mtr{F} \\
     \vdots & \ddots & \vdots \\
     \omega_{N-1}^0 \mtr{F} & \ldots & \omega_{N-1}^{N-1} \mtr{F} \\
     \end{array}
      \right).
\end{equation}

Let us consider a function of two angles $u(\phi,\psi)$. Sampling $u$ on a square grid
of $\phi_j = j\,2 \pi/N$, and $\psi_k = k\,2 \pi/N$, where $j,k = 0, \ldots,N-1$,
we create a vector $\vv{u} \in \mathbb{R}^{N^2}$, whose $p$-th element is
\begin{equation}\label{u:p}
    \vv{u}[p] = u(\phi_j,\psi_k), \quad p = j\,N+k.
\end{equation}

The direct discrete Fourier transform (DFT) of $u$ is
the vector $\vv{\hat{u}} \in \mathbb{R}^{N^2}$ resulting from the matrix-vector product
\begin{equation}\label{DFT}
    \vv{\hat{u}}  = \frac{1}{N^2}\,\mtr{F}_2 \vv{u}.
\end{equation}
The inverse DFT is provided by the complex conjugate  $\mtr{F}_2^\ast$ with the property
\begin{equation}
     \mtr{F}_2^\ast \mtr{F}_2 = \mtr{F}_2 \mtr{F}_2^\ast = N^2\,\mtr{I},
\end{equation}
so that
\begin{equation} \label{DFTI}
    \mtr{F}_2^\ast\,\vv{\hat{u}} =  \frac{1}{N^2}  \mtr{F}^\ast \mtr{F} \vv{u} = \vv{u},
\end{equation}
explaining the necessity of the $N^{-2}$ factor in Eq.~(\ref{DFT}).

We can consider DFT as the coefficients of a trigonometric polynomial
\begin{equation}\label{trig}
    u \approx \sum_{j=-N'}^{N'} \sum_{k=-N'}^{N'}
    u_{jk} \mathrm{e}^{\mathrm{i} \left( j \phi + k \psi \right)},
    \quad N' = \lfloor N/2 \rfloor,
\end{equation}
where $\lfloor~\rfloor$ is the ``floor'' rounding down operator.  Introducing
\begin{equation}\label{Z}
    Z_N(q) = \left\{
\begin{array}{lcl}
  q & \mbox{for} & q \leqslant \lfloor \mbox{$\frac{1}{2}$} N \rfloor, \\
  q-N & \mbox{for} & q >  \lfloor\mbox{$\frac{1}{2}$} N  \rfloor,
\end{array}
    \right.
\end{equation}
we can identify
\begin{eqnarray}\label{tr:coef}
    \vv{\hat{u}}[q_1\,N+q_2] & = & u_{jk}, \\
    j & = & Z_N(q_1), \\
    k & = & Z_N(q_2),
\end{eqnarray}
with the indices $q_1,q_2 = 0, \ldots, N-1$.
Strictly speaking, for even $N$ the Nyquist terms
with $|j| = N/2$ or $|k| = N/2$ require a special treatment and an extra factor $1/2$ or $1/4$, but their influence on the final solution
is practically so marginal, that we do not pay attention to this problem.

\section{PPM with regolith layer}
\label{Regol}

The thermal model presented in Sect.~\ref{Thermal} can be easily extended to cover a case when
an asteroid is treated as monolithic core covered by a regolith layer of thickness $h$.
Let us assume conductivity $K$ and thermal diffusivity $\kappa$ for the regolith layer $0 \leqslant \zeta < h$,
and  $K_\mathrm{c}$, $\kappa_\mathrm{c}$ for the core $\zeta \geqslant h$.
Then, Eq. (\ref{Helm:1}) is replaced by two sets
\begin{eqnarray}\label{Helm:r}
 \left( \vv{\hat{T}}[p] \right)'' - \mathrm{i}\,\beta_{p}\, \vv{\hat{T}}[p] & = & 0, \\
 \label{Helm:c}
 \left( \vv{\hat{T}}_\mathrm{c}[p] \right)'' - \mathrm{i}\,w^2\,\beta_{p}\, \vv{\hat{T}}_\mathrm{c}[p] & = & 0,
\end{eqnarray}
where $p = 0,\ldots,N^2-1$, and $w^2$ is the ratio of thermal diffusivities
\begin{equation}\label{nu}
    w = \sqrt{\frac{\kappa}{\kappa_\mathrm{c}}},
\end{equation}
introduced to use a single parameter $\beta_p$ defined for the surface layer according to Eq.~(\ref{beta}).\footnote{Our $w$ is the same as $\xi_1$ of \citet{VB:99}.}

Solutions $\vv{\hat{T}}_\mathrm{c}[p]$ are subject to the Neumann condition at infinity, hence, similarly to
(\ref{T:sol}) they read
\begin{equation}\label{Tc:sol}
 \vv{\hat{T}}_\mathrm{c}[p] = C_p\, \exp\left[ - \left(1+\mathrm{sgn}(\beta_p)\,\mathrm{i} \right)\,\sqrt{\frac{w\,|\beta_p|}{2}}\,\zeta \right],
\end{equation}
except for the special case $\vv{\hat{T}}_\mathrm{c}[0]=C_0$.
For the regolith layer, however, the asymptotic condition does not apply, so it takes a more general form $\vv{\hat{T}}[p]$
\begin{eqnarray}
 \vv{\hat{T}}[p] & = &
 A_p\, \exp\left[ - \left(1+\mathrm{sgn}(\beta_p)\,\mathrm{i}  \right)\,\sqrt{\frac{|\beta_p|}{2}}\,\zeta \right] \nonumber \\
& & + B_p\, \exp\left[  \left(1+\mathrm{sgn}(\beta_p)\,\mathrm{i}  \right)\,\sqrt{\frac{|\beta_p|}{2}}\,\zeta \right],
\label{Tr:sol}
\end{eqnarray}
with the boundary condition (\ref{bcv:2}). As usually, the special case $\vv{\hat{T}}[0] = A_0$ applies.

Both solutions should satisfy continuity requirements at $\zeta = h$, i.e. additional Dirichlet conditions
\begin{equation}\label{Diri}
    \left( \vv{\hat{T}}_\mathrm{c}[p] - \vv{\hat{T}}[p] \right)_{\zeta=h} = 0,
\end{equation}
and Neumann conditions
\begin{equation}\label{Neu}
    \left( \vv{\hat{T}}'_\mathrm{c}[p] - \vv{\hat{T}}'[p] \right)_{\zeta=h} = 0.
\end{equation}
Thanks to them, we can express $B_p$ in terms of $A_p$ alone, and so we obtain the logarithmic derivative
that generalizes (\ref{T:rat}) for the surface temperature for $p \neq 0$
\begin{equation}\label{T:rat:2}
    \gamma_p =  \frac{\vv{\hat{T}'}[p]}{\vv{\hat{T}}[p]} =
    - \left(1+  \mathrm{sgn}(\beta_p)\, \mathrm{i} \right)\,
    \frac{w-1 + (w+1) H_p}{1-w + (w+1) H_p}\,\sqrt{\frac{|\beta_p|}{2}},
\end{equation}
where
\begin{equation}\label{Hp}
    H_p = \mathrm{e}^{\left(1+ \mathrm{sgn}(\beta_p)\,\mathrm{i}\right)\, h\,\sqrt{2\,|\beta_p|}},
\end{equation}
and the special case with $p=0$ is $\gamma_0 =  0$.

The formula (\ref{T:rat:2}) uses no assumptions about the depth $h$, but if we postulate a thin regolith layer,
the  terms linear in $h$ are simply
\begin{equation}\label{T:rat:lin}
    \gamma_p  \approx
    - \left(1+  \mathrm{sgn}(\beta_p)\, \mathrm{i} \right)\,
   w\,\sqrt{\frac{|\beta_p|}{2}}  + \mathrm{i}\,\beta_p\,\left(w^2-1\right)\,h.
\end{equation}
Recalling that $w^2 \beta_p$ is the ratio of insolation frequency to the thermal diffusivity of the core, we see that $\gamma_p$
is dominated by the properties of the inner part of asteroid, although entering the matrix $\mtr{B}$ it will be multiplied by
the surface conductivity of the regolith layer $K$.

Going even further and neglecting $h$ in the approximation (\ref{T:rat:lin}), we obtain the simplest recipe for the regolith covered objects:
use the physical parameters of the core to compute $\gamma_p$, and the regolith parameters in the rest of the algorithm.
Of course, it would be wise to verify the quality of this approximation by comparing at least the values of
$\gamma_1$ computed according to (\ref{T:rat:lin}) with $h=0$, with the ones obtained using a more exact formula.

Concluding this section we emphasize that even when the physical properties of an object vary with the depth $\zeta$, our
conclusion about the independence of $\langle\dot{\omega}\rangle$ on conductivity remain true.

\section{Spherical segments as a 1D model with finite depth}

\label{Finde}

Suppose that an object is starlike, i.e. each surface element can be connected with the centre of mass using a straight segment that does not intersect
other surface elements. In these circumstances, we can consider the Laplacian operator in spherical coordinates,
and the reduction to 1D amounts to neglecting non-radial part of $\Delta$, so that Fourier equation (\ref{Four:2}) becomes
\begin{equation}\label{Four:C}
 \frac{\kappa}{r^2}\,\frac{\partial~}{\partial r} \left( r^2\,\frac{\partial T}{\partial r}\right) =  \frac{\partial T}{\partial t},
\end{equation}
where $0 \leqslant r \leqslant R$. An elementary substitution
\begin{equation}\label{udef}
    u = \frac{r}{R}\,T,
\end{equation}
reduces Eq.~(\ref{Four:C}) to the form similar to (\ref{Four:3})
\begin{equation}\label{Four:C:1}
 \kappa\,\frac{\partial^2 u}{\partial r^2} =  \frac{\partial u}{\partial t},
\end{equation}
so the associated Helmholz equation is the same as (\ref{Helm:1})
\begin{equation}\label{Helm:C:1}
 \frac{\mathrm{d}^2 \vv{\hat{u}}[p]}{\mathrm{d}r^2} - \mathrm{i}\,\beta_p\, \vv{\hat{u}}[p] = 0, \qquad p = 0,\ldots,N^2-1.
\end{equation}
Note, that at the surface, where $r=R$, we have $u=T$, allowing us to use the algorithm of Sect.~\ref{Thermal} directly,
without even changing the symbols. But first we have to redefine the coefficients $\gamma_p$ to account for new boundary conditions.

Instead of asymptotic Neumann conditions at $\zeta \rightarrow \infty$, typical for the plane-parallel case, we now have the Dirichlet condition
$u_p(r=0) = 0$, satisfied automatically when $T$ at the origin is finite. On the surface, where $u=T$, the energy balance
(\ref{bc:1}) holds true, but now the gradient of $T$, reduced to the radial derivative, is not the same as the derivative of $u$, because
\begin{equation}\label{deru}
    \left[ \frac{\mathrm{d} T}{\mathrm{d} r} \right]_{r=R} = \frac{\mathrm{d} u}{\mathrm{d} r} - \frac{u}{R}.
\end{equation}
Thus, instead of (\ref{Tprim:pp}) we use
\begin{equation}\label{QC}
    Q = K \delta \left( \frac{u}{R}  - \frac{\mathrm{d} u}{\mathrm{d} r} \right),
\end{equation}
where $\delta$ is the cosine of the angle between the radius vector and the outward normal vector of the current surface element.
As we see, there are two major differences in the conduction treatment between the PPM and the spherical segment approach: the dependence on
the local radius $R$, and the deviation of the gradient from the normal direction.

Similarly to the PPM model, we solve Eq.~(\ref{Helm:C:1}) as a harmonic oscillator with imaginary frequency, but this time
a different boundary condition gives us at the surface $r=R$
\begin{equation}
  \frac{1}{ \vv{\hat{u}}[p]}  \frac{\mathrm{d} \vv{\hat{u}}[p]}{\mathrm{d} r} =
 \sqrt{\mathrm{i}\,\beta_p}\,\coth\left( R\,\sqrt{\mathrm{i}\,\beta_p} \right),
\end{equation}
and the ratios $\gamma_p$ in matrix $\mtr{B}$ should be replaced by
\begin{equation}
 \gamma_p  =  \delta \left( \, \frac{1}{R} -  \frac{1}{ \vv{\hat{u}}[p]}  \frac{\mathrm{d} \vv{\hat{u}}[p]}{\mathrm{d} r} \right),
\end{equation}
when $p \neq 0$.
However, when $p=0$, we still have
\begin{equation}\label{g0:C}
    \gamma_0 = 0,
\end{equation}
so the YORP effect in rotation rate $\omega$ remains insensitive to conductivity, similarly to the PPM case.

Taking the outward normal parallel to the radius, i.e. fixing $\delta=1$, and taking the limit at infinite $R$, when $h_p \rightarrow 1$, $t_p \rightarrow 0$,
we recover the plane-parallel model with $\gamma_0=0$, and remaining $\gamma_p$ given by Eq.~(\ref{T:rat}).

\label{lastpage}
\end{document}